\documentclass[reprint, showpacs,
preprintnumbers,
amsmath,amssymb,
aps, floatfix,
superscriptaddress,
prd, nofootinbib]{revtex4-1}

\usepackage{float}
\usepackage{graphicx}
\usepackage{dcolumn}
\usepackage{bm}
\usepackage{hyperref}
\usepackage[mathlines]{lineno}
\usepackage{xcolor}
\usepackage{import}
\usepackage{booktabs}

\usepackage[capitalise, nameinlink]{cleveref}
\usepackage{makecell}
\usepackage{apptools}

\begin{document}
\title{First Constraints on WIMP-Nucleon Effective Field Theory Couplings in an Extended Energy Region From LUX-ZEPLIN}

\author{J.~Aalbers}
\affiliation{SLAC National Accelerator Laboratory, Menlo Park, CA 94025-7015, USA}
\affiliation{Kavli Institute for Particle Astrophysics and Cosmology, Stanford University, Stanford, CA  94305-4085 USA}

\author{D.S.~Akerib}
\affiliation{SLAC National Accelerator Laboratory, Menlo Park, CA 94025-7015, USA}
\affiliation{Kavli Institute for Particle Astrophysics and Cosmology, Stanford University, Stanford, CA  94305-4085 USA}

\author{A.K.~Al Musalhi}
\affiliation{University College London (UCL), Department of Physics and Astronomy, London WC1E 6BT, UK}

\author{F.~Alder}
\affiliation{University College London (UCL), Department of Physics and Astronomy, London WC1E 6BT, UK}

\author{C.S.~Amarasinghe}
\email{amarascs@umich.edu}
\affiliation{University of Michigan, Randall Laboratory of Physics, Ann Arbor, MI 48109-1040, USA}

\author{A.~Ames}
\affiliation{SLAC National Accelerator Laboratory, Menlo Park, CA 94025-7015, USA}
\affiliation{Kavli Institute for Particle Astrophysics and Cosmology, Stanford University, Stanford, CA  94305-4085 USA}

\author{T.J.~Anderson}
\affiliation{SLAC National Accelerator Laboratory, Menlo Park, CA 94025-7015, USA}
\affiliation{Kavli Institute for Particle Astrophysics and Cosmology, Stanford University, Stanford, CA  94305-4085 USA}

\author{N.~Angelides}
\affiliation{Imperial College London, Physics Department, Blackett Laboratory, London SW7 2AZ, UK}

\author{H.M.~Ara\'{u}jo}
\affiliation{Imperial College London, Physics Department, Blackett Laboratory, London SW7 2AZ, UK}

\author{J.E.~Armstrong}
\affiliation{University of Maryland, Department of Physics, College Park, MD 20742-4111, USA}

% ...

\collaboration{LZ Collaboration}
\date{\today}

\begin{abstract}
Following the first science results of the LUX-ZEPLIN (LZ) experiment, a dual-phase xenon time projection chamber operating from the Sanford Underground Research Facility in Lead, South Dakota, USA, we report the initial limits on a model-independent non-relativistic effective field theory describing the complete set of possible interactions of a weakly interacting massive particle (WIMP) with a nucleon. These results utilize the same 5.5 t fiducial mass and 60 live days of exposure collected for the LZ spin-independent and spin-dependent analyses while extending the upper limit of the energy region of interest by a factor of 7.5 to 270~keV$_\text{nr}$. 
No significant excess in this high energy region is observed. 
Using a profile-likelihood ratio analysis, we report 90\% confidence level exclusion limits on the coupling of each individual non-relativistic WIMP-nucleon operator for both elastic and inelastic interactions in the isoscalar and isovector bases.
\end{abstract}
    
\maketitle


\section{\label{sec:introduction}Introduction}
Current and next-generation dark matter (DM) direct detection experiments searching for weakly interacting massive particles (WIMPs) will reach unprecedented levels of exposure during their operational lifetimes, allowing them to probe new parameter space for WIMPs, as demonstrated by the first LZ~\cite{LZ:SR1WS_2022} and XENONnT~\cite{XenonNT:WS_2023} spin-independent (SI) and spin-dependent (SD) WIMP-nucleon interaction limits. 
Both the SI and SD interaction models assume that the momentum dependence of the interaction is suppressed due to the non-relativistic velocity of the WIMP in the nucleus rest frame, hence allowing for the use of a zero momentum cross-section~\cite{JUNGMAN:Supersymmetric_dark_matter}. 
However, this may be an oversimplification of the dynamics of the interaction as, despite being small, the velocity is still non-zero.
By considering interactions with momentum dependence, it is possible to increase the potential for the discovery of WIMP-nucleon interactions by direct detection experiments.
\par
The rate of a more generalized interaction may depend on the momentum transfer; if the momentum-independent components are suppressed, as in the interaction of a nucleon and DM with a magnetic dipole moment, the dominant component will be momentum-dependent.
When considering all possible interactions in the nucleon frame, the momentum may be non-negligible and is certainly non-negligible for WIMP-parton interactions~\cite{Hisano:2017jmz}.
Given the variety of possible WIMP-nucleon interactions, the most comprehensive approach is to consider the WIMP-nucleon interaction in a model-independent way. 
Since the non-gravitational interactions of DM are unknown, it is compelling to describe the interaction through an effective field theory (EFT) that captures a significant amount of the possible physics~\cite{georgi:EFT}.
Constraints on a comprehensive set of WIMP-nucleon interactions generated using an EFT~\cite{Fan_2010, Fitzpatrick:EFT}, have previously been established by PandaX-II~\cite{PandaX2:SD_EFT_2019}, LUX~\cite{LUX:EFTR4_2021}, XENON100~\cite{Xenon100:EFT_2017}, DEAP-3600~\cite{DEAP:eft_2020}, Darkside-50~\cite{DarkSide-50:eft_2020} and PICO-60~\cite{PICO:photonM_2022}.
This report further constrains the coefficients within this EFT by applying a profile likelihood ratio analysis to the first science data of the LZ experiment.
An extended energy window is used to increase the coverage of the signal model profiles, which may produce significant interaction rates at energies beyond that of a typical SI or SD WIMP search. 
As such, the calibrations, efficiencies, and relevant backgrounds are reassessed for this wider energy range.

\subsection{\label{subsec:theory}Non-Relativistic Effective Field Theory} 
A full derivation of the WIMP-nucleon non-relativistic effective field theory (NREFT), which describes all possible interactions between WIMPs and a target nucleus, has been developed by Fan \textit{et al.}~\cite{Fan_2010} and Fitzpatrick \textit{et al.}~\cite{Fitzpatrick:EFT}.
In an NREFT, one expands the effective Lagrangian in powers of the momentum transferred in the WIMP-nucleus interaction. 
The cutoff scale for this theory is dictated by the maximum momentum transfer, around 200 MeV, set by the reduced mass of the WIMP-nucleus system and escape velocity of the WIMP~\cite{Anand:MathematicaEFT}.
The pion degrees of freedom are integrated out of this EFT, restricting its validity to energies below the mass of the pion~\cite{Fitzpatrick:EFT}.
While this NREFT reliably describes all possible interactions at these energies, in order to match it to an ultraviolet theory of dark matter, the QCD dynamics that were ignored during the creation of the NREFT have to be brought back. 
Examples of doing this are described in Refs.~\cite{Bishara_2017,Hoferichter_2018,Xenon1t:eft_2022} and experimental limits on WIMP-pion interactions are described in Ref.~\cite{Xenon1t:wimp-pion-eft-2019}.

\par
In the Fitzpatrick \textit{et al.} NREFT~\cite{Fitzpatrick:EFT}, the corresponding complete interaction Lagrangian can be written in terms of linear combinations of effective operators, each multiplied by a coefficient. 
These operators can be expressed in terms of WIMP-nucleon couplings, which are proportional to low-energy constants (parameters such as the WIMP axial-vector or scalar coupling constant) in both the effective Lagrangian and the nuclear form factor. 
The nuclear form factor describes the spatial distribution of the nuclear density and is calculated using models that consider the structure and dynamics of the target nucleus.
The NREFT is thus a powerful tool for interpreting experimental data in a model-independent way. 
Furthermore, the power counting~\cite{BUCHALLA201480:EFT} scheme of NREFT allows for a systematic and controlled expansion of the scattering cross-section in terms of the momentum transfer, which is essential for calculating the sensitivity of direct detection experiments to WIMP dark matter.
The derivation of the NREFT introduces both momentum-dependent and velocity-dependent operators to describe all possible mechanisms for the WIMP-nucleon interaction. 
By enforcing momentum conservation and Galilean invariance \cite{Fitzpatrick:EFT}, the operators can be reduced to a basis of four Hermitian quantities:
\begin{equation}
i \frac{\vec{q}}{m_N}, \quad \vec{v}^\perp \equiv \vec{v} + \frac{\vec{q}}{2\mu}, \quad \vec{S}_{\chi}, \quad \vec{S}_{N},
\label{eq:basis}
\end{equation}
\noindent where $\vec{q}$ is the momentum transferred from the WIMP to the nucleon, $m_N$ is the nucleon mass, $\vec{v}^\perp$ is the component of the relative velocity between the WIMP and the nucleon that is perpendicular to the momentum transfer, $\vec{S_\chi}$ is the spin of the WIMP, and $\vec{S_N}$ is the spin of the relevant nucleon. 
By forming linear combinations of the Hermitian quantities, up to second-order in $\vec{q}$, a total of  fifteen independent and dimensionless NREFT operators can be constructed~\cite{Anand:MathematicaEFT}:

\begin{equation}
\begin{array}{l}
{\mathcal{O}_{1}=1_\chi 1_N},\quad
{\mathcal{O}_{2}=\left(v^{\perp}\right)^{2}}, \quad
{\mathcal{O}_{3}=i \vec{S}_{N} \cdot\left(\frac{\vec{q}}{m_N} \times \vec{v}^{\perp}\right)}, \\ 
{\mathcal{O}_{4}=\vec{S}_{\chi} \cdot \vec{S}_{N}}, \quad
{\mathcal{O}_{5}=i \vec{S}_{\chi} \cdot(\frac{\vec{q}}{m_N} \times \vec{v}^{\perp})}, \\ 
{\mathcal{O}_{6}=\left(\vec{S}_{\chi} \cdot \frac{\vec{q}}{m_N}\right)\left(\vec{S}_{N} \cdot \frac{\vec{q}}{m_N}\right)}, \quad
{\mathcal{O}_{7}=\vec{S}_{N} \cdot \vec{v}^{\perp}}, \\
{\mathcal{O}_{8}=\vec{S}_{\chi} \cdot \vec{v}^{\perp}}, \quad
{\mathcal{O}_{9}=i \vec{S}_{\chi} \cdot\left(\vec{S}_{N} \times \frac{\vec{q}}{m_N}\right)}, \\
{\mathcal{O}_{10}=i \vec{S}_{N} \cdot \frac{\vec{q}}{m_N}}, \quad 
{\mathcal{O}_{11}=i \vec{S}_{\chi} \cdot \frac{\vec{q}}{m_N}}, \\

{\mathcal{O}_{12} =\vec{S}_{\chi} \cdot\left(\vec{S}_{N} \times \vec{v}^{\perp}\right)}, \quad
{\mathcal{O}_{13} =i\left(\vec{S}_{\chi} \cdot \vec{v}^{\perp}\right)\left(\vec{S}_{N} \cdot \frac{\vec{q}}{m_N}\right)}, \\
{\mathcal{O}_{14} =i\left(\vec{S}_{\chi} \cdot \frac{\vec{q}}{m_N}\right)\left(\vec{S}_{N} \cdot \vec{v}^{\perp}\right)}, \\
{\mathcal{O}_{15} =-\left(\vec{S}_{\chi} \cdot \frac{\vec{q}}{m_N}\right)\left(\left(\vec{S}_{N} \times \vec{v}^{\perp}\right) \cdot \frac{\vec{q}}{m_N}\right)} .
\end{array}
\label{eq:operators}
\end{equation}

As only non-relativistic interactions are considered in this analysis, any operator with a quadratic or greater dependency on $v^{\perp}$ is not included (namely $\mathcal{O}_2$). 
This analysis is conducted in the isoscalar ($s$) and isovector ($v$) bases as opposed to the proton and neutron bases used in some previous analyses. 
It can be argued that proton and neutron are much more natural bases to produce WIMP-nucleon scattering constraints, as this is what the WIMP interacts with.
However, by exploiting the fact that isospin is a good quantum number in nuclear systems, it is possible to decompose the fundamental constraints on the operators when comparing experiments with varying target nuclei.
An additional benefit to using the isoscalar and isovector bases is that it is possible to compare to the limits set on the elastic and inelastic WIMP-nucleon NREFT operators by the XENON collaboration~\cite{Xenon100:EFT_2017}, as well as the inelastic limits set by LUX~\cite{LUX:EFTR4_2021}.
\par
This report also considers inelastic scatters where the masses of the incoming and outgoing DM particles can differ, allowing for the WIMP to transition into a more massive state during the scattering interaction~\cite{Smith_2001}. 
This mass difference, denoted by $\delta_m \equiv m_{\chi,\mathrm{out}} - m_{\chi,\mathrm{in}}$, results in a minimum required recoil energy to produce an inelastic interaction. This in turn leads to a suppressed recoil rate at lower energies and an observed signal that is concentrated at higher energies. 
In some cases, where elastic scattering is suppressed, inelastic transitions between WIMP states become the primary method of interaction~\cite{Han_1997, Hall_1998}.
A slight modification of the Hermitian basis vectors is required to study the case of inelastic WIMP-nucleon interactions. 
In elastic interactions, energy conservation results in $\vec{v}^{\perp} \cdot \vec{q} = 0$. To preserve energy conservation in inelastic interactions where there is a nonzero mass splitting, $\delta_m$, the following condition must be satisfied~\cite{Barello_2014}:
\begin{equation}\label{eq:massSplitting} 
\delta_m + \vec{v} \cdot \vec{q} +\frac{|\vec{q}|^2}{2\mu_N }= 0, 
\end{equation}
where $\mu_N$ is the reduced mass of the WIMP-nucleon system~\cite{Barello_2014}. 
This requirement is incorporated into the Hermitian basis vectors by replacing the perpendicular velocity in \autoref{eq:basis} with
\begin{equation}\label{eq:idmVelocity} 
\vec{v}^\perp_{inel} \equiv \vec{v} + \frac{\vec{q}}{2\mu_N} + \frac{\delta_m}{|\vec{q}|^2}\vec{q} =v^\perp + \frac{\delta_m}{|\vec{q}|^2}\vec{q}, 
\end{equation}
where $v^\perp$ is the perpendicular velocity for elastic scattering. 
A similar replacement is made for each operator, $\mathcal{O}_i$. 
This report considers mass splitting values in the 0-250~keV range, which are well-motivated in many WIMP models~\cite{Smith_2001, Barello_2014}.
\par
\begin{figure*}[hbt!]
    \centering
    \includegraphics[trim={8 5 5 5},clip, width=\textwidth]{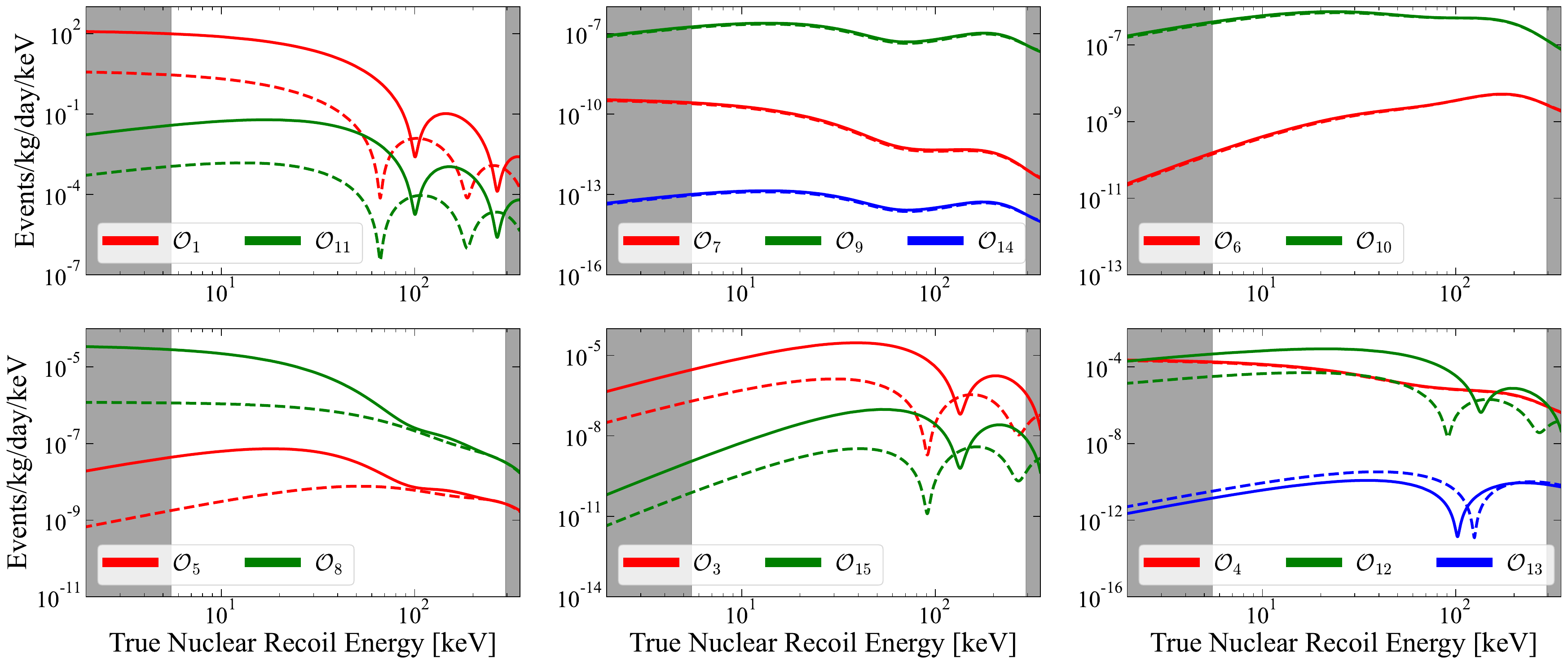}
    \caption{
    Differential recoil spectra for the fourteen non-relativistic NREFT WIMP-nucleon operators for a 1000~GeV/c$^2$ WIMP in both the isoscalar (solid line) and the isovector (dashed line) bases.
    The operators are categorized by their nuclear responses (as defined in Ref.~\cite{Anand:MathematicaEFT}): $M$-only (top-left), $\Sigma^{'}$-only (top-center), $\Sigma^{''}$-only (top-right), $M$ and $\Delta$ (bottom-left), $\Sigma^{'}$ and $\Phi^{''}$ (bottom-center) and all other responses (bottom-right).
    The spectra were generated with a coupling strength of unity, excluding the possibility of interference terms. 
    The shaded gray regions indicate the energies at which the detection efficiency is below 50\% after all data analysis cuts have been applied (see~\autoref{fig:acceptances}).}
    \label{fig:recoils}
\end{figure*}

Before experimental limits can be translated into bounds on the operator coefficients of \autoref{eq:operators}, it is necessary to calculate nuclear response functions for DM elastic scattering. 
It is common practice to use the full-basis shell-model calculations of each contributing Xe isotope using the GCN5082 interaction when determining these responses~\cite{MENENDEZ2009139}. 
Recently, these calculations have been revised, leading to a much-improved determination of the one-body nuclear density matrices~\cite{haxton_unpublished}.
In this analysis, the updated density matrices were incorporated into DMFormfactor-v6~\cite{Anand:MathematicaEFT}, a software package for calculating the NREFT dark matter-nucleus scattering. 
These response functions were subsequently integrated into WimPyDD, a framework for modeling the interaction of dark matter with atomic nuclei in direct detection experiments~\cite{Jeong_2022}. 
This modified version of WimPyDD was validated against both DMFormfactor-v6 and the published interaction Lagrangian recoil spectra by PandaX and Haxton~\cite{PandaX2:SD_EFT_2019}.
\autoref{fig:recoils} shows the differential rate spectra, generated using the modified WimPyDD, for each non-relativistic operator considered in this analysis, assuming a WIMP mass of 1000~GeV/c$^2$.

\section{\label{sec:lz_data_selection} Detector, Data and Selection}
\subsection{\label{subsec:selection}The LZ Detector} 
The LZ experiment is housed in the Davis Campus of the Sanford Underground Research Facility, 4850 (4300~m.w.e) below Lead, South Dakota, USA, which provides an attenuation of the cosmic muon flux by a factor of 10$^6$~\cite{LZ:Experiment_2020,LZ:TDR_2017}.
LZ utilizes a low-background, dual-phase time projection chamber (TPC) which contains 7~tonnes of liquid xenon (LXe) in the active volume~\cite{LZ:Experiment_2020,LZ:TDR_2017}. 
The cylindrical TPC, with both a height and diameter of approximately 1.5~m, is equipped with two arrays consisting of 251 and 243 R11410–22 photomultiplier tubes (PMTs) at the top and bottom of the detector, respectively, to detect scintillation light. 

Energy depositions in the LXe volume are detected by collection of the scintillation produced from the initial interaction (S1), as well as from the electroluminescent light from extracted electrons (S2).
Electrons are collected due to the presence of electric fields applied across the detector volume: a 192~V/cm drift field and a 7.3~kV/cm gas electroluminescence field. 
The S1 and S2 signals are reported in units of photons detected (phd), accounting for the double photon emission effect, which can occur when VUV light is incident on a PMT. A requirement that scintillation light is detected in at least three PMTs (three-fold coincidence) is used when defining an S1 in the data.
\par
Additionally, the S2:S1 ratio differs for electronic recoils (ERs) and nuclear recoils (NRs), providing discrimination power between different interaction types.
This is illustrated in \autoref{fig:calibrations} using LZ calibration datasets.
The position-corrected S1 and S2 observables are typically used (S1$c$ and S2$c$, respectively),  to improve the energy resolution and discrimination between ERs and NRs. 
Position-corrected observables also allow for the use of linear scaling factors, $g_1$ and $g_2$, to correlate the S1$c$ and S2$c$ signals, respectively, to the original number of photons and electrons produced ($n_{ph}$ and $n_e$): 
\begin{equation}
    \text{S1}c = g_1\langle n_{ph} \rangle; \hspace{1cm} \text{S2}c = g_2 \langle n_e \rangle. 
\label{eq:S1S2scale}
\end{equation}
\par
To reject internal and external backgrounds, the LZ experiment includes two veto detectors: a xenon ``Skin" veto surrounding the active mass, which is outfitted with 93 1-inch and 38 2-inch PMTs; and a near-hermetic ``outer detector" (OD) consisting of acrylic tanks containing 17~tonnes of gadolinium-loaded liquid scintillator (0.1\% by mass). 
The skin is designed to identify multiple scattering interactions entering or exiting the TPC, while the outer detector is designed to capture and identify neutrons that may scatter in the TPC. 
The entire system is housed in a tank that is filled with 238~tonnes of ultra-pure water, which provides shielding from ambient radioactive backgrounds emitted by the cavern rock~\cite{LZ:Experiment_2020, LZ:TDR_2017}.
120 8-inch PMTs are situated around the walls of this tank to observe any light produced in the OD.
\par
A 60 live-day exposure using a 5.5~$\pm$~0.2 tonne fiducial volume (FV) was collected between December 2021 and May 2022. 
Using events from a dataset with S1$c$ between 3~and~80~phd, uncorrected S2~$>$~600~phd and log$_{10}$(S2$c$)~$<$~5, LZ set world-leading constraints on SI WIMP-nucleon interactions for masses greater than 9~GeV/c$^2$, 
with the most stringent limit excluding cross-sections greater than 9.2~$\times$~10$^{-48}$~cm$^2$ for 36~GeV/c$^2$ WIMPs~\cite{LZ:SR1WS_2022}. 
This report extends the region of interest (ROI) of the same dataset to include S1$c$ signals between 3~and~600~phd to give sensitivity to NREFT interactions with their most significant rate contribution outside the typical search window used when considering SI and SD interactions. 
The maximum S2$c$ considered in this analysis is set at log$_{10}$(S2$c$)~=~4.5 to remove ER backgrounds far from the NR signal region, given that leakage from ERs becomes less significant at higher energies (see \autoref{fig:calibrations}).
The lower bound on uncorrected S2 is maintained at 600~phd.

\subsection{\label{subsec:cal_and_data}Calibrations and Data Selections}
\par
As described in Ref.~\cite{LZ:SR1WS_2022}, the ER and NR responses were measured using dedicated \textit{in-situ} calibrations with tritiated methane (0--18.6~keV ERs) and D-D fusion neutrons (0--74~keV NRs). 
The detector and model response parameters from NEST 2.3.7 (Noble Element Simulation Technique)~\cite{NEST:paper_2022, NEST:paper_2023} were tuned to the ER and NR calibration data in order to reproduce the observed data. 
This tuning was used along with constraints from the energy reconstruction of several monoenergetic peaks from background and calibration sources to find the S1$c$ and S2$c$ scaling factors defined in \autoref{eq:S1S2scale}: $g_1$~=~0.114~$\pm$~0.002~phd/photon and $g_2$~=~47.1~$\pm$~1.1~phd/electron. 

Extending the ER and NR response models through the extended energy ROI is one of the key challenges in performing an NREFT analysis. For this, $\beta$ emissions from ${}^{212}$Pb following an injection of ${}^{220}$Rn provide a calibration of the ER response through the 3~--~600~phd NREFT search window. 
Reproducing the ${}^{212}$Pb response does not require altering $g_1$ or $g_2$ from Ref~\cite{LZ:SR1WS_2022}; however, NEST underestimates the observed ER band widths above 100~phd, worsening at higher energies ($\mathcal{O}$(10\%) disagreement at 600~phd). 
To account for this, the functionality in NEST for energy-dependent smearing of pulse areas is utilized, similar to the methods reported in Ref.~\cite{LUX:ER_modelling_2020}. 
This results in proper reproduction of the ER response in and beyond the NREFT ROI, allowing for accurate characterization of ER leakage from $\beta$ backgrounds into the NR signal region.
\par
\begin{figure}[ht]
    \centering
    \includegraphics[trim={8 5 5 5},clip, width=0.44\textwidth]{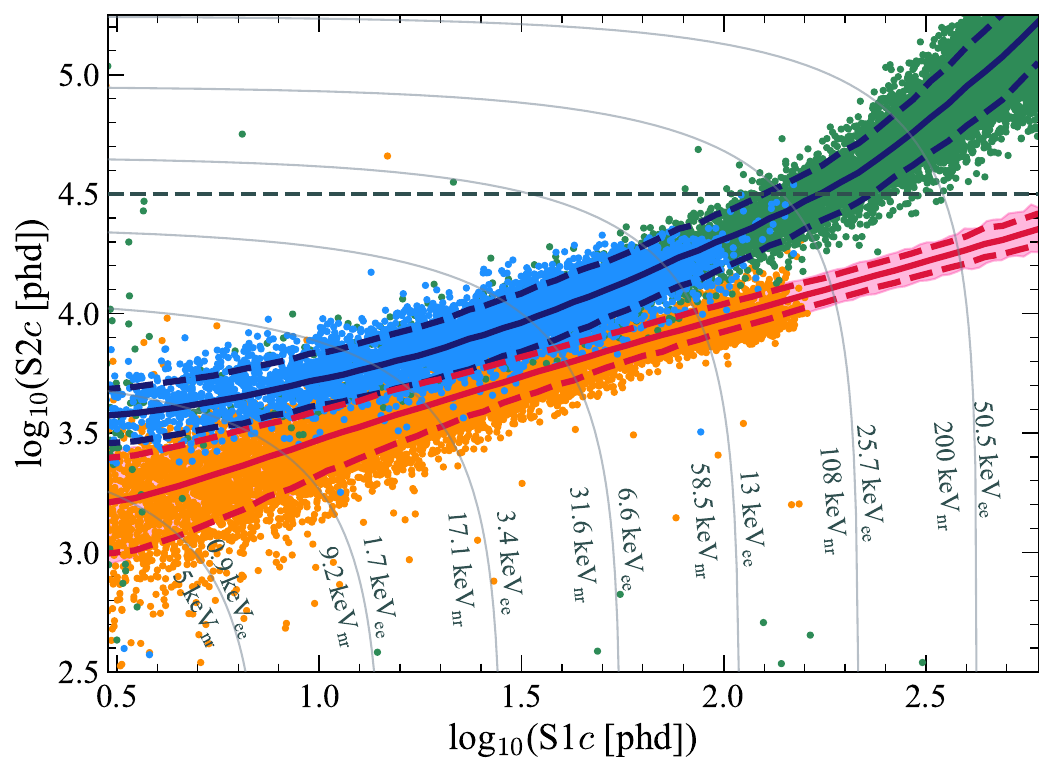}
    \caption{Calibration events in \{log$_{10}$(S1$c$), log$_{10}$(S2$c$)\} space from tritium (light blue), D-D neutrons (orange), and ${}^{212}$Pb (green) in and above the ROI.
    The mean ER and NR responses from NEST for flat energy spectra are shown in dark blue and red, respectively; the dashed lines indicate the 10\%~--~90\% percentiles of the expected response.
    Additionally, the region highlighted in pink denotes the shift in the 10\%~--~90\% percentiles when considering the $\pm1\sigma$ uncertainty in the NEST mean NR response beyond the D-D endpoint used for this study.
    Gray contours are lines of constant reconstructed energy, labeled for both ER and NR interactions. 
    The gray dashed horizontal line denotes the upper log$_{10}$(S2$c$) bound used in this analysis, however, the calibration events in this region were used to constrain the ER response model. 
    }
    \label{fig:calibrations}
\end{figure}

The NR response for LZ is only directly calibrated to approximately 80~keV$_{\rm{nr}}$\footnote{For energy units, ``keV" is used for true recoil energies, while reserving ``keV$_{\rm{ee}}$" and ``keV$_{\rm{nr}}$" for energy values reconstructed from measured quantities and assuming either an ER or NR interaction, respectively.} using D-D with supplementary AmLi neutron calibrations. 
However, the NR models in NEST are based on a collection of all LXe light and charge yield measurements available in the existing literature, and the highest energy measurements included in NEST are the 330~keV yields from AmBe reported by Sorensen \textit{et al.}~\cite{Sorensen_2011}. 
Therefore this analysis relies on extrapolating the NEST NR response beyond the \textit{in-situ} LZ calibrations, specifically using recent \textit{ex-situ} measurements of the NR response from D-T neutrons which provide information on the NR yields up to 426~keV~\cite{DT_calib}. 
Ref.~\cite{DT_calib} suggests that the NEST models may be overestimating both the light and charge yields at higher energies beyond the D-D endpoint. 
Using those D-T measurements, the uncertainty in the NEST NR models beyond the \textit{in-situ} calibration yields is constrained. 
However, when accounting for the reduction in total quanta, the change in the mean NR \{log$_{10}$(S1$c$), log$_{10}$(S2$c$)\} response is calculated to be $<$1\%. 
The NEST models for the $\beta$ ER and NR response and NR response uncertainty used in this analysis are compared to LZ calibration data and shown in \autoref{fig:calibrations}.
The impact from this uncertainty on WIMP sensitivity is tested by altering the NR response between -1$\sigma$ and +3$\sigma$ uncertainty. 
Because the discrimination between ERs and NRs significantly improves as energy increases, the NR response uncertainty has a negligible impact on the results of this analysis ($<$10\% on the final WIMP sensitivity, for any combination of mass and operator).

\par
A series of data selection criteria are used to remove accidental coincidences of isolated S1s and S2s (``accidentals'') from true single scatters at an efficiency greater than 99.5\%. 
Sources of isolated S1 pulses are particle interactions in charge-insensitive regions of the TPC, Cherenkov and fluorescent light in detector materials, and dark-current pile-up. 
Isolated S2 pulses can be generated from radioactivity and electron emission from the cathode or gate electrodes, particle interactions in the gas phase or the liquid above the gate electrode, and delayed drift electron signals~\cite{LUX:2020vbj}.  
The criteria for removing accidentals are tuned using side band events and are only applied to the search dataset after being finalized. 

The accidental removal criteria are the same as the methods reported in Ref.~\cite{LZ:SR1WS_2022}, using relationships between pulse and event based quantities (such as drift time, the ratio between light collected in either the top or bottom PMT arrays, pulse width, the timing of PMT hits within the pulse, and hit pattern of the photons in the PMT arrays), and targeted individual sources of isolated S1 and S2 pulses by comparing to the expected single scatter behavior.
The efficiency of the data selection criteria beyond the lower-energy ROI reported in Ref.~\cite{LZ:SR1WS_2022} is evaluated using tritium and $^{220}$Rn data for cuts targeting S1 pulses values and the combination of tritium and AmLi data for those targeting S2 pulses. 

\begin{figure}[t]
    \centering
    \includegraphics[trim={8 5 5 5},clip, width=0.44\textwidth]{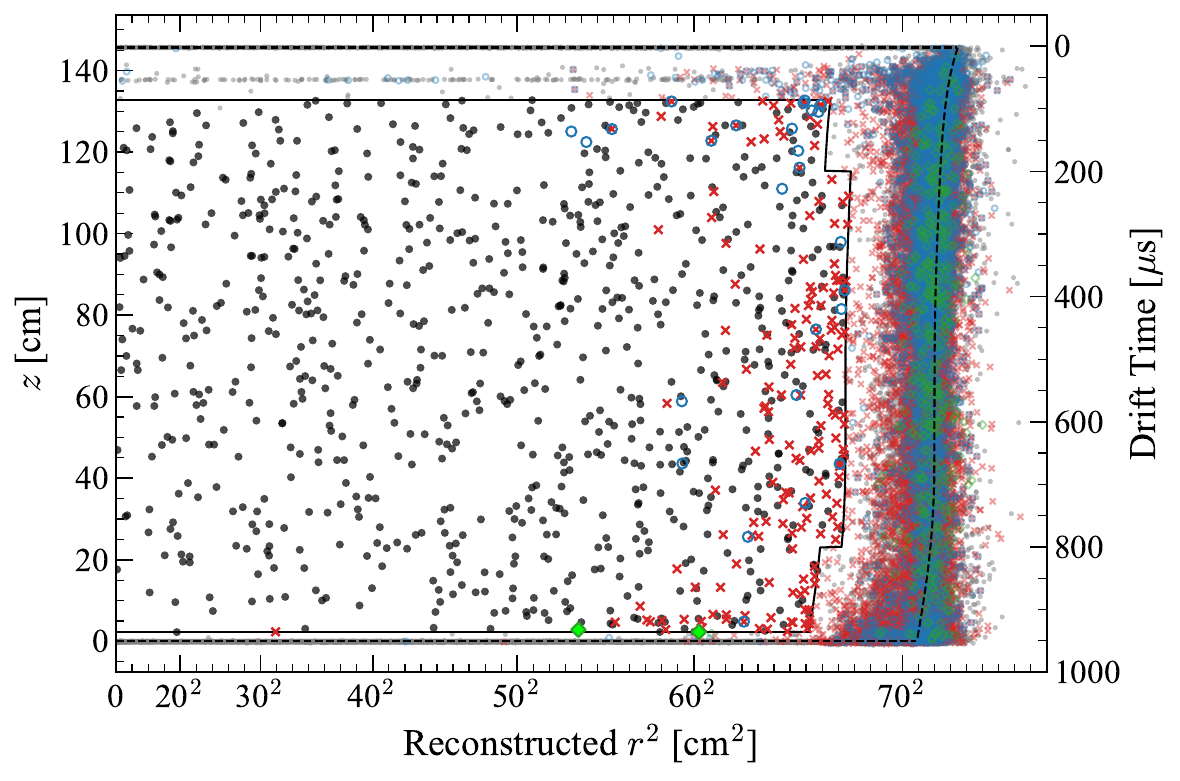}
    \caption{Data in reconstructed $r^2$ and $z$ after all analysis cuts. 
    Black (gray) points show the data inside (outside) the fiducial volume after all cuts and vetoes have been applied.
    Red crosses and blue circles show events vetoed by a prompt Skin and OD signal, respectively.
    Solid green diamonds indicate the events removed by the $\gamma$-X BDT cut after all other cuts have been applied. 
    Hollow diamonds indicate events outside the FV classified as $\gamma$-X.
    The solid line shows the fiducial volume definition, and the dashed line shows the extent of the active TPC.
    }
    \label{fig:r2_drift}
\end{figure}

Following a similar approach as the LUX NREFT analysis~\cite{LUX:EFTR4_2021}, this analysis implements a boosted decision tree (BDT) to identify and remove $\gamma$-X events, which are the interactions of multiply scattering $\gamma$-rays classified as single scatters due to one or more scatters occurring in a region of the TPC from where charge cannot be collected~\cite{LUX:EFTR4_2021}.
This leads to an attenuated S2 signal relative to the observed S1, increasing the ER leakage into the signal region.
Two events are identified as $\gamma$-X and removed from the LZ NREFT search data, consistent with the model expectation of 1.6 events, all from the bottom of the FV. 
\autoref{fig:r2_drift} shows the spatial distribution of the events passing all analysis criteria, highlighting the events removed by the $\gamma$-X classifier and the LXe Skin and OD veto systems.
The NR acceptance of the BDT was calculated using simulated data to be $99.950 \pm 0.002$\% within the FV, and was validated using D-D and AmLi neutron calibration data. 
A fraction of events is classified as $\gamma$-X outside the FV, where the BDT performance is degraded due to the noisiness of S1 signals near the reflective TPC wall. 
Further details of the $\gamma$-X event topology, modeling procedure, and BDT results are discussed in \autoref{subsec:mssi}. 
\par
\begin{figure}[t]
    \centering
    \includegraphics[trim={5 5 5 5},clip, width=0.44\textwidth]{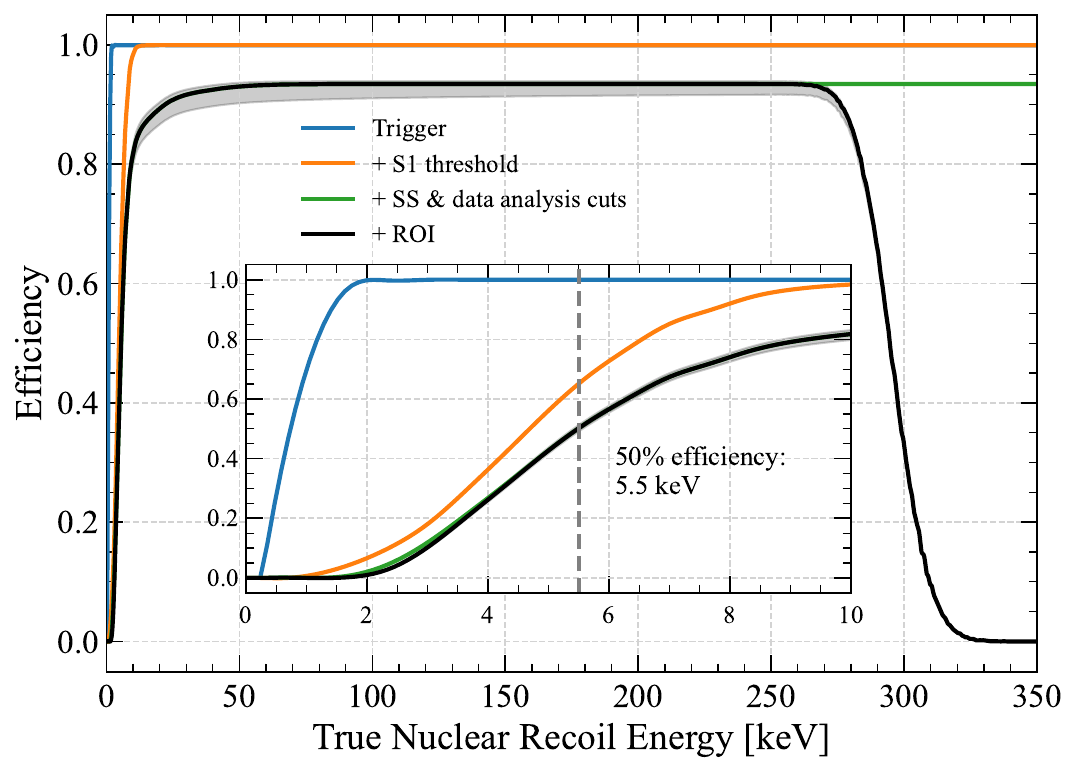}
\caption{Signal efficiency as a function of the NR energy from the trigger (blue), the three-fold coincidence and \>3~phd threshold on S1$c$ (orange), the single-scatter (SS) reconstruction and analysis cuts (green), and the search ROI in S1$c$ and S2$c$ (black). 
The low energy behavior is shown in the inset, where the dotted line at 5.5~keV$_{\text{nr}}$ indicates the nuclear recoil energy at which the efficiencies equal 50\%.
The uncertainty on the detection efficiency (gray region) was assessed with ${}^3$H, ${}^{220}$Rn, and AmLi calibration data. 
}
    \label{fig:acceptances}
\end{figure}

The final efficiency of the data selection criteria, evaluated with ${}^3$H, ${}^{220}$Rn, and AmLi calibration data and including trigger and event reconstruction efficiency, is shown in \autoref{fig:acceptances} as a function of true NR energy.

\section{\label{sec:model_and_stats}Modeling}
To simulate the background and signal components in the observable space, the BACCARAT package based on GEANT4 \cite{LZ:simulations_2021, ALLISON2016} is utilized, along with a bespoke simulation of the LZ detector response, which is fine-tuned using the NEST detector model. 
As part of this methodology, the uncertainties associated with the background components are included as constraint terms in a combined fit of the background model to the data.

\begin{figure}
    \centering
    \includegraphics[trim={8 5 5 5},clip, width=0.44\textwidth]{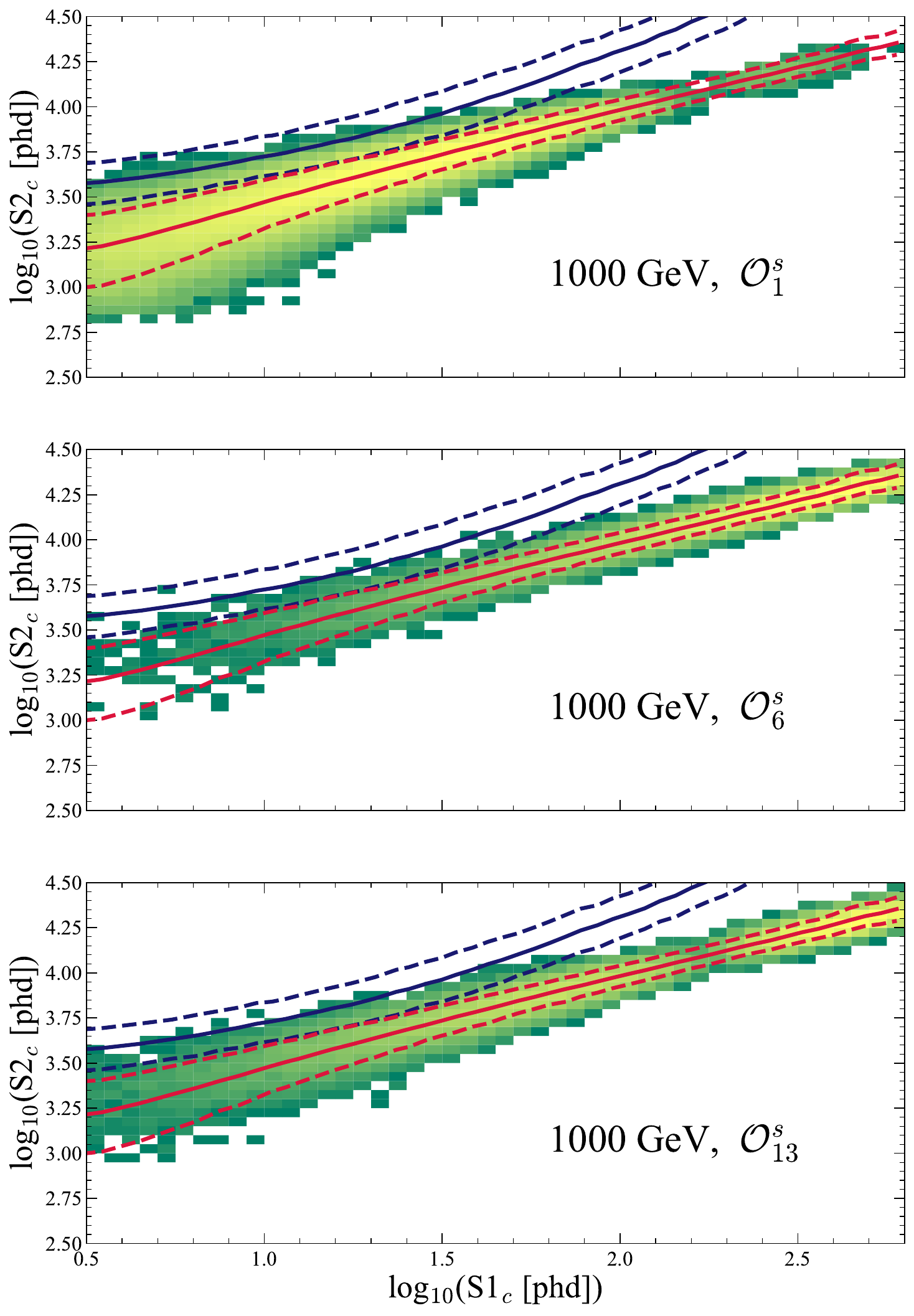}
    \caption{Example \{log$_{10}$(S1$c$), log$_{10}$(S2$c$)\} distributions for a 1000~GeV/c$^2$ WIMP-nucleon isoscalar interaction for the momentum-independent operator $\mathcal{O}_1$ (top), and momentum-dependent operators $\mathcal{O}_6$ (middle) and $\mathcal{O}_{13}$ (bottom). 
    Red and blue lines denote the flat ER and NR response regions as described in \autoref{fig:calibrations}.
    Each pane shows the distribution for 100,000 WIMP nuclear recoils.
    }
    \label{fig:signal-PDF}
\end{figure}

\subsection{\label{subsec:signal}Signals}
\par
As stated in \autoref{sec:introduction}, the recoil spectra for each operator-mass combination are generated using a modified version of WimPyDD. 
This analysis considers both isoscalar and isovector interactions for WIMP masses between 10 and 4000~GeV/c$^2$.
As several operators have a dependence on the spin of the target nucleus, the response from each Xe isotope, weighted by its natural abundance, is incorporated into the overall response of the target.  
In this analysis, the isospin representation follows the methodology outlined by Anand \textit{et al.}~\cite{Anand:MathematicaEFT}, previously used in Refs.~\cite{PandaX2:SD_EFT_2019, Xenon100:EFT_2017} where
\begin{equation}
    c^{s} = \frac{1}{2} (c^{p} + c^{n}); \hspace{1cm} c^v = \frac{1}{2}(c^p - c^n).
\end{equation}
$c^n$ and $c^p$ denote the coupling coefficient for interactions with neutrons and protons, respectively, and $c^s$ and $c^v$ represent isoscalar or isovector interactions, respectively.
The normalization chosen here differs from some previous searches such as Ref.~\cite{LUX:EFTR4_2021}, where $c^{s} = (c^p + c^n)$, which will produce a factor 4 difference from the rates here.
\par
Following the convention set in Ref.~\cite{DM_parameters:BAXTER2021_Conventions}, the Standard Halo Model is used to describe the WIMP velocity distribution $f(v)$ with $\vec{v}_\circledast$~=~(11.1, 12.2, 7.3)~km/s (solar peculiar velocity)~\cite{Schoenrich:Local_kinematics}, $\vec{v}_0$~=~(0, 238, 0)~km/s (local standard of rest velocity)~\cite{DM_parameters:galaxy_context_rest_velocity_1, DM_parameters:galaxy_context_rest_velocity_2} and $v_{\text{esc}}$~=~544~km/s (galactic escape speed)~\cite{DM_parameters:RAVE_survey_escape_velocity}. 
The local DM density $\rho_0$ is taken as 0.3~GeV/cm$^3$ \cite{DM_parameters:LEWIN199687_DM_density}. 
\par
Each operator is considered as an independent interaction channel, where only contributions from that operator with no interference have been considered. 
In practice, this requires setting the a priori of the coefficient of the given operator to $1/m_v$, and all other coefficients to 0. 
The quantity $m_v$ is introduced to ensure the coefficients have dimensions of energy$^{-2}$ and is set to the Higgs vacuum expectation value of 246.2~GeV/c$^2$~\cite{Fitzpatrick:EFT}. 
This is a natural value for the normalization as it allows the coefficients to be expressed in terms of the Standard Model weak interaction mass scale.
The normalization is required due to the decision by Anand \textit{et al.} to normalize the spinors by 4$m_N m_\chi$, which allows for a dimensionless representation of the operators. 
By considering the case of a single operator $\mathcal{O}_i$ dominating the interaction, the differential recoil rate scales with the square of $c_i^N$, where $N \in \{s, v\}$, such that
\begin{equation}\label{eq:spectrumReduced}
     \frac{\mathrm{d} R_i^N}{\mathrm{d} E_{R}}=\frac{(c_{i}^{N})^2 \rho_{0}}{32 \pi m_{\chi}^{3} m_{N}^{2}} \int_{v>v_\text{min}} \frac{f(\vec{v})}{v} F_{i, i}^{\left(N, N\right)}\left(v^2, q^{2}\right) \mathrm{d}{v},
\end{equation}
where $v_\text{min} = q(m_\chi + m_N) / (2 m_\chi m_N)$ is the minimum velocity that produces a recoil energy $E_R$, and $F_{i,i}^{(N, N)}$ is a form factor that depends on the nuclear physics. 
A sample of the signal distributions in the $\{$log$_{10}$(S1$c$), log$_{10}$(S2$c)\}$ observable space is shown in \autoref{fig:signal-PDF}.

\subsection{\label{subsec:bg}Backgrounds}
\par
The background model used in this analysis consists of 11 components.
\autoref{tab:fit_result} lists the expected and fitted number of events for each component.
Ref.~\cite{LZ:sr1backgrounds_2022} contains a complete discussion of the backgrounds in LZ for the data-taking period used for this analysis.
Ref.~\cite{LZ:SR1WS_2022} describes most of these background sources in detail, and the expectation for some backgrounds is mostly unchanged (such as ${}^{37}$Ar, ${}^{8}$B coherent neutrino-nuclear scattering, detector neutrons, and accidental coincidence of isolated S1s and S2s). 
The expected contributions from continuous ER sources -- namely the ``Flat ER" sum of ${}^{222}$Rn, ${}^{220}$Rn, and ${}^{85}$Kr, as well as the ERs from solar neutrinos and double-$\beta$ decays from ${}^{136}$Xe -- is increased from Ref.~\cite{LZ:SR1WS_2022} due to the extended energy window of this analysis. 
Except for increasing the energy range, these models are unchanged from the previous analysis. 
The ${}^{127}$Xe and ${}^{124}$Xe models are expanded in this analysis, as the ${}^{127}$Xe K-shell electron captures and ${}^{124}$Xe KL, KM, and KN double electron capture signals are reconstructed partially within the search ROI\footnote{A dedicated NEST model is used to model electron capture ERs, as the mean light and charge yields per unit energy for these interactions has been shown to differ from $\beta$ interactions~\cite{XELDA_Lshell_EC}.}.

Two other sources of background are treated uniquely for this analysis: ${}^{125}$I electron captures and Compton scattering $\gamma$-rays from trace levels of ${}^{40}$K, ${}^{60}$Co, ${}^{232}$Th, and ${}^{238}$U in the detector components~\cite{LZ:radioactivity_and_cleanliness_2020} as well as ${}^{40}$K, ${}^{232}$Th, and ${}^{238}$U from the cavern walls \cite{LZ:cavern_gamma_2020}. 

${}^{125}$I is introduced into the TPC via neutron activation of ${}^{124}$Xe into ${}^{125}$Xe and its subsequent decay. 
With a 59.4~d half-life, ${}^{125}$Xe produces unresolved multiple scatters with a 35.5~keV $\gamma$-ray in addition to an electron capture X-ray. 
The combined $\gamma$+L (40.4~keV) and $\gamma$+M (36.5~keV) decays contribute events into the search window of this analysis. 
The $\gamma$+K decay is outside of this search window and was used to infer the rate of ${}^{125}$I. 

Compton scatters from detector components were treated similarly to the Flat ER contributions in Ref.~\cite{LZ:SR1WS_2022}. 
However, the rate of these backgrounds increases with energy as higher energy decays have longer mean free paths in LXe~\cite{berger1998xcom}. 
The longer mean free paths increase the probability of multiple scatters in the fiducial volume. 
Due to finite resolution, some of the events are not separable from single scatters, causing an increased rate of events to deviate from a standard ER single scatter response at higher energies.
This is further obscured by the unique detector pathologies near detector components, such as poor light collection efficiency and field fringing, increasing the deviation.
Therefore, detector-based ERs are separated from other ER sources for the background fitting procedure in this analysis. 
These effects are typically subdominant in terms of ER leakage into the signal region since the ER and NR bands diverge at these energies unless a significant portion of the energy is deposited in a region from where the ionization signal cannot be collected. 
These $\gamma$-X events predominantly occur near the cathode and TPC walls. 
Because they are a unique background to high-energy searches and can be reconstructed near and below the NR signal region, they are considered separately in this analysis and described in the following section.

\begin{figure}[ht]
    \centering
    \includegraphics[trim={8 5 5 5},clip, width=0.44\textwidth]{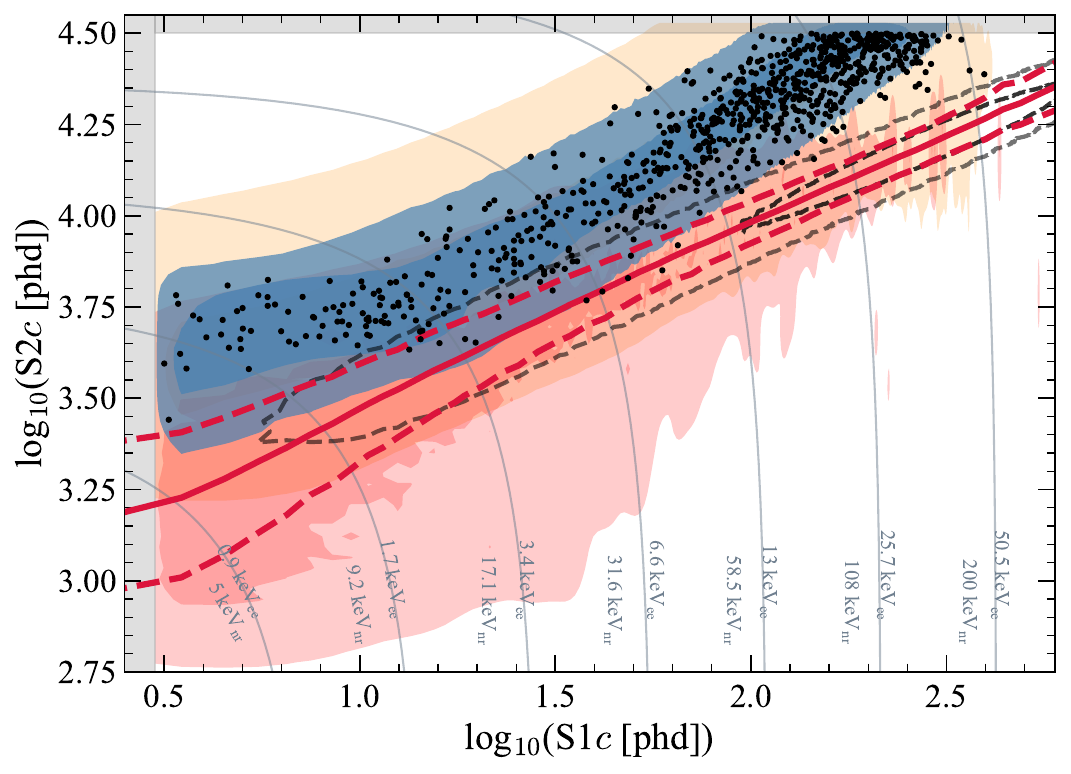}
\caption{The final high energy WIMP-search data after all cuts in \{log$_{10}$(S1$c$), log$_{10}$(S2$c$)\} space. 
    The contours that enclose 1$\sigma$ (dark) and 2.5$\sigma$ (light) regions represent the following models: 
    the shaded red region indicates the detector neutrons, the shaded orange region indicate the detector ERs, the blue region is the combined representation of all other ER models (${}^{214}$Pb, ${}^{212}$Pb, ${}^{85}$Kr, ${}^{37}$Ar,  ${}^{125}$I, ${}^{124}$Xe, ${}^{127}$Xe, ${}^{136}$Xe, and $\nu$ ER) and the black dashed lines show a 1000 GeV/c$^2$ $\mathcal{O}_6$ isoscalar signal model. 
    The solid red line corresponds to the NR median, while the red dotted lines represent the 10~--~90\% percentiles of the expected response.
    The model contours are produced with a linear scale for S1$_c$ prior to being cast into log and take into account all the efficiencies used in the analysis. 
    Contours of constant recoil energy have been included as thin gray lines. Grayed regions at the left and top of the plot indicate parameter space outside the energy ROI.
}

    \label{fig:data-ROI}
\end{figure}

\hyperref[fig:data-ROI]{Figure~\ref*{fig:data-ROI}} shows the \{log$_{10}$(S1$c$), log$_{10}$(S2$c$\} distribution of the 835 events which pass all selections, along with contours representing a 1000 GeV/c$^2$ $\mathcal{O}_6$ isoscalar signal model (representative of signal models that peak at non-zero energy), and the background model.

\begin{table}[ht]
    \centering
    \caption{Expected and fitted numbers of events from the listed sources in the 60~d~$\times$~5.5~t exposure.
    The middle column shows the predicted number of events with uncertainties as described in the text.
    These uncertainties are used as constraints in a combined fit of the background model.
    The fit result is shown in the right column. ``Flat ER" represents the combination of ${}^{214}$Pb, ${}^{212}$Pb, and ${}^{85}$Kr mixed in the LXe, while ``Detector ER" represents electron recoils originating from radiogenic decays in detector materials.
    Both ${}^{37}$Ar and the detector neutrons have non-gaussian constraints and are totaled separately.
    Values with a fit result of zero are set to have no lower uncertainty.
    }
    \begin{ruledtabular}
    \begin{tabular}{ccc}
        Source                   & Expected Events      & Fit Result        \\ \hline
        Flat ER                  & 517.4~$\pm$~82.8     & 574.7~$\pm$~30.2  \\
        Detector ER              & 18.4~$\pm$~9.2       & 22.3~$\pm$~8.1    \\
        $\nu$ ER                 & 55.3~$\pm$~5.5       & 55.5~$\pm$~5.5    \\
        ${}^{124}$Xe             & 8.2~$\pm$~2.0        & 8.7~$\pm$~2.0     \\
        ${}^{127}$Xe             & 20.5~$\pm$~1.8       & 20.8~$\pm$~1.8    \\
        ${}^{136}$Xe             & 55.1~$\pm$~11.6      & 58.2~$\pm$~11.2   \\
        ${}^{125}$I              & 30.1~$\pm$~15.6      & 34.2~$\pm$~8.9    \\
        ${}^{8}$B CE$\nu$NS      & 0.14~$\pm$~0.01      & 0.14~$\pm$~0.01   \\
        Accidentals              & 1.3~$\pm$~0.3        & 1.3~$\pm$~0.03    \\ \hline
        Subtotal                 & 706~$\pm$~86         & 775~$\pm$~34 \\ \hline
        ${}^{37}$Ar              & [0, 288]             & 49.5~$\pm$~9.4    \\
        Detector neutrons        & 0.0$^{+0.5}$         & 0.0$^{+1.8}$      \\ \hline
        Total                    & -                    & 825~$\pm$~36  
    \end{tabular}
    \end{ruledtabular}
    \label{tab:fit_result} 
\end{table}

\subsubsection{\label{subsec:mssi}$\gamma$-X Events}
\par
A $\gamma$-X event occurs when a $\gamma$-ray scatters multiple times in the TPC, but at least one scatter occurs in a region where electrons cannot be extracted (such as below the cathode electrode or in regions with significant electric field distortions).
This leads to missing S2 pulses observed in the interaction, allowing these multiple scatters to be erroneously classified as single scatters with lower S2:S1 ratios than typical ER events, potentially mimicking an NR event. 
These pathologies can be ignored in a traditional SI WIMP search; the S2-prohibitive regions are typically spatially distant from the FV boundaries, so it becomes unlikely that a $\gamma$-X event can occur in the FV with an S1 signal small enough to be observed in a standard WIMP low energy ROI~\cite{LZ:SR1WS_2022}. 
However, in the extended energy window of an NREFT search, $\gamma$-X events become a significant NR-like background.
\par
In LZ, the locations primarily responsible for the production of $\gamma$-X events are the reverse field region (RFR) below the cathode, and in close proximity to the walls of the TPC.
Both regions have electric fields that do not allow the complete collection of charge.
In the case of the RFR, electrons drift down instead of up.
In the vicinity of the TPC wall, non-uniformities in the electric field direct drifting electrons to the wall, leading to the depletion of the charge signal.
The most significant source of $\gamma$-X events below S1$c$~$=600$~phd is the electron capture decay of $^{127}$Xe near the cathode, where the two-step de-excitation produces an X-ray photon (0.2-32.2~keV) and a $\gamma$-ray photon with energy $\mathcal{O}$(100~keV).
This same decay near the TPC wall and decays of rare-earth impurities in the cathode grid wires have minor contributions to the overall rate of $\gamma$-X events. 
\par
To generate $\gamma$-X events from $^{127}$Xe and cathodic origins, we use a custom Monte Carlo simulation for $\gamma$-ray propagation through a realistic LZ geometry.
By incorporating the measured activities of ${}^{127}$Xe and radiogenic impurities in detector components, this simulation calculates the three-dimensional propagation of $\gamma$-rays from a given starting position, using the expected energy-dependent mean free path and Compton scattering cross-sections in LXe~\cite{weinberg1995quantum, berger1998xcom}. 
Light and charge yields are calculated using the photo-absorption and Compton scattering ER models from NEST.
S1 position dependence and individual PMT channel areas are calculated using light collection maps generated with BACCARAT~\cite{LZ:simulations_2021}. 
Additionally, field magnitudes and electron drift paths throughout the TPC are calculated using maps generated from dedicated finite-element simulations~\cite{fenics}, providing information about the regions along the cathode and wall where an S2 signal would escape detection.
The simulations include the potential wells at the wall created by TPC field shaping rings that trap electrons, reducing the size of the S2 pulse.
However, the rate of $\gamma$-X events at the wall is far lower than that at the cathode, given the much larger volume of the RFR. 
\par
A data-driven number of expected $\gamma$-X events was calculated using the $\gamma$-X model and the measured rate of $^{127}$Xe, the dominant source of the background.
The model was used to obtain the ratio of $\gamma$-X to the clean 203~keV peak from $^{127}$Xe, which was then scaled by the size of the 203~keV peak in the observed dataset to yield an expectation of 1.6 $\pm$ 0.3 $\gamma$-X events in the ROI and FV. 
The rate of observed $\gamma$-X candidates rises in accordance with the model outside the FV as the edge of the TPC boundary is approached. 
\par
The $\gamma$-X simulations are used to train a multi-class BDT in order to classify events as ER, NR, $^{127}$Xe RFR $\gamma$-X, $^{127}$Xe wall $\gamma$-X, or cathode RFR $\gamma$-X.
The BDT is provided with a seven-dimensional simulated dataset with the following features: 
\begin{itemize}
\vspace{-0.25cm}
    \item[--] position-corrected S1 pulse area,
    \vspace{-0.25cm}
\item[--] position-corrected S2 pulse area,
   \vspace{-0.25cm}
 \item[--] radial reconstructed position,
    \vspace{-0.25cm}
\item[--] vertical position (drift time),
    \vspace{-0.25cm}
\item[--] ``cluster size",  the dispersion of S1 light collected in the bottom PMT array,
    \vspace{-0.25cm}
\item[--] ``max channel fraction", the ratio of light observed in the brightest bottom array PMT to the total bottom array S1 area,
    \vspace{-0.25cm}
\item[--] the ratio of light observed in the top and bottom PMT arrays.
\end{itemize}
The final three S1 pulse-based features may be exploited to differentiate $\gamma$-X events because each of the multiple scatters contribute to the S1 pulse. 
None of the seven features individually can provide sufficient $\gamma$-X rejection power. 
The information contained in the correlations of the hit pattern features with pulse areas and event positions, however, may be harnessed by a multivariate tool such as a BDT for the clean removal of $\gamma$-X events.  
\par
The accuracy of true single scatter and $\gamma$-X identification is assessed by cross-validating BDT predictions for ten non-overlapping datasets.
In this procedure, the entire simulated dataset is split into ten equally-sized portions (80,000 events), and each portion is taken as the validation set for a BDT that is trained on the remaining nine portions.
The parameters and classification thresholds of the ten BDTs are identical to those of the final BDT deployed on the EFT search data. 
The general performance on the validation datasets is summarized in \autoref{table: reduced confusion matrix cuts} with errors given by the standard deviations of predicted counts across the ten BDTs. 
A high averaged single scatter acceptance is seen within the FV and ROI, which remains high even at the boundaries of the FV. 
\begin{table}
    \begin{center}
    \caption{Confusion matrix after FV and ROI cuts, averaged over ten BDTs (each evaluated on a different dataset), showing the correct identification rate (\%) across the two classes on the diagonal and the misclassification rate on the off-diagonals.}
        \begin{ruledtabular}
            \begin{tabular}{c|cc}
                 & True SS & True $\gamma$-X \\
                \hline
                Predicted SS & $99.997 \pm 0.005$ & $0.4 \pm 1.2$ \\
                Predicted $\gamma$-X & $0.003 \pm 0.005$ & $99.6 \pm 1.2$ \\
            \end{tabular}
        \end{ruledtabular}
    \end{center}
    \label{table: reduced confusion matrix cuts}
\end{table}
To validate the high acceptance of simulated single scatters in observed data, the BDT is deployed on $^{220}$Rn calibration data in the S1c~$<$~1000~phd range.
Under the most conservative assumption that all 25,000 events originating from the $^{220}$Rn decay chain are true single scatters, the single scatter acceptance rate of the BDT is $99.92$\%. 
The BDT is also deployed on DD and AmLi neutron calibration data to verify NR acceptance. 
No DD events are removed by the BDT, and only AmLi events below the NR band that are indicative of multiply scattering neutrons were removed (multiply scattering neutron events have the same S1 hit pattern characteristics of $\gamma$-X events). 
Finally, the $\gamma$-X rejection rate in the 1000~phd~$<$~S1$c$~$<$~2000~phd side band of the WIMP search data is tested.
This side band has a significant number of $\gamma$-X events from the 375~keV $\gamma$-ray of the $^{127}$Xe electron capture decay.
The BDT removes 73\% of the 180 $\gamma$-X events on and below the NR band.
The 27\% $\gamma$-X misclassification rate is attributed to differences in the data and the detector response model (used to train the BDT), which is not tuned or validated in this S1 range.

The assessment using calibration data shows that the BDT has an acceptance larger than 99.9\% for ER and NR single scatter events throughout the FV.  
A lower bound on the $\gamma$-X rejection rate of 73\% is obtained from a side band above the ROI, with the true rejection rate in the ROI expected to approach the value from simulations (99.6\%) at lower energies due to the better match between data and the detector response model. 
Two events are removed by the BDT in the search dataset, consistent with the expectation of 1.6 $\gamma$-X events found using the measured activities and the rate at which they produce $\gamma$-X in the custom Monte Carlo simulation.
This consistency, in addition to the high $\gamma$-X rejection rate of the BDT classifier, removes the need to incorporate a $\gamma$-X model in the fits to the observed data. 

\section{\label{sec:results}Results}
No significant evidence of an excess is found among the NREFT operators in the isoscalar and isovector bases for both elastic and inelastic DM in any of the models tested. 
Comparisons of the reconstructed energy distributions between the data and the background model using unbinned Kolmogorov-Smirnov (Anderson-Darling) tests yield p values of 0.392 (0.25), showing consistency between the data and the background-only scenario.
Upper limits on the DM coupling strengths for each NREFT interaction are presented in \autoref{subsec:elastic} for elastic scatters and \cref{subsec:inelastic} for DM upscattering to a heavier state.

The upper limits are obtained by defining an extended unbinned profile likelihood statistic in log$_{10}$(S2$c$)-S1$c$ space, which is used to construct two-sided bounds at the 90\% confidence level following Refs.~\cite{LZ:SR1WS_2022, DM_parameters:BAXTER2021_Conventions}. 
The resulting coupling strength limits are cast in the dimensionless form $(c_i^N \times m_v^2)^2$, where $m_v = 246.2$~GeV/c$^2$ is the Higgs vacuum expectation value \cite{Anand:MathematicaEFT}. 

\subsection{Elastic}\label{subsec:elastic}
\par
\begin{figure*}
    \includegraphics[width=0.5\columnwidth]{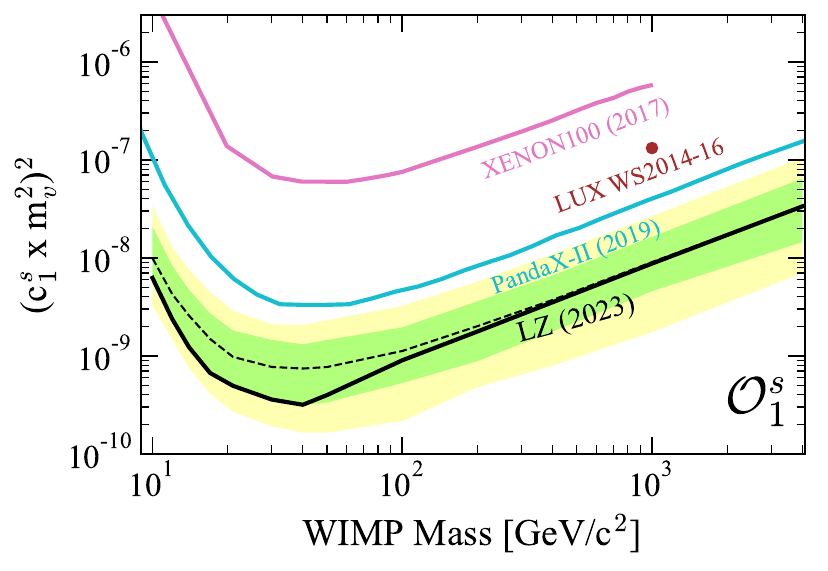}
    \includegraphics[width=0.5\columnwidth]{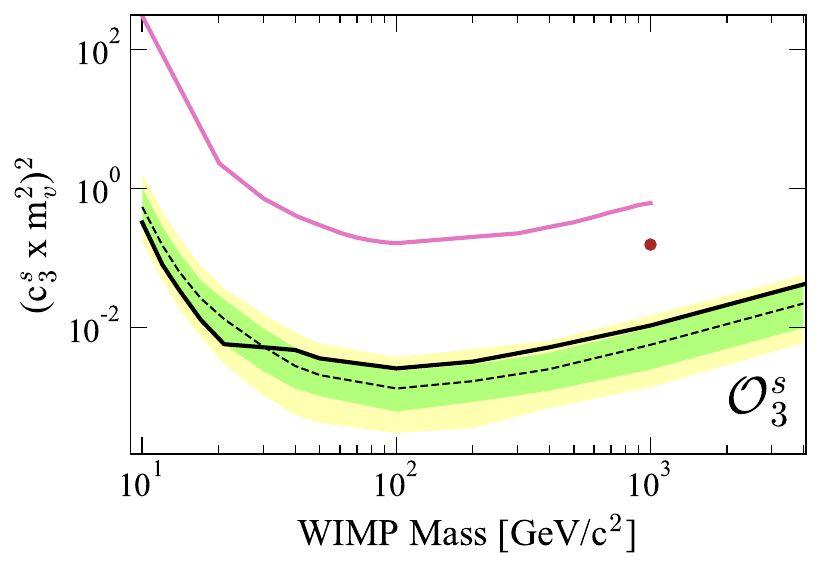}
    \includegraphics[width=0.5\columnwidth]{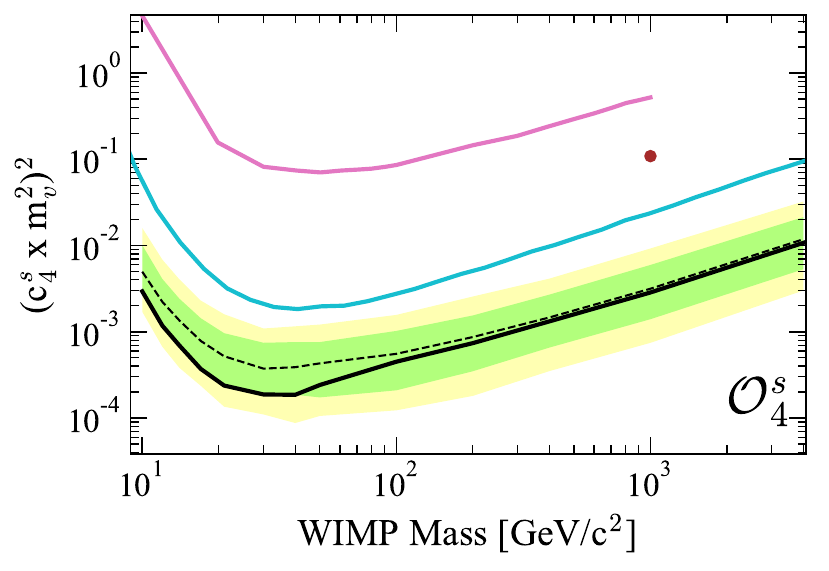}
    \includegraphics[width=0.5\columnwidth]{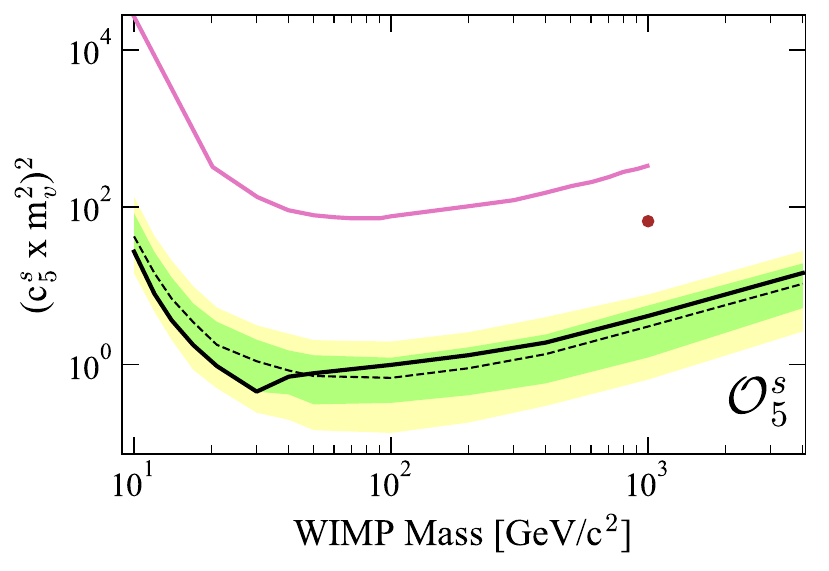}
    \includegraphics[width=0.5\columnwidth]{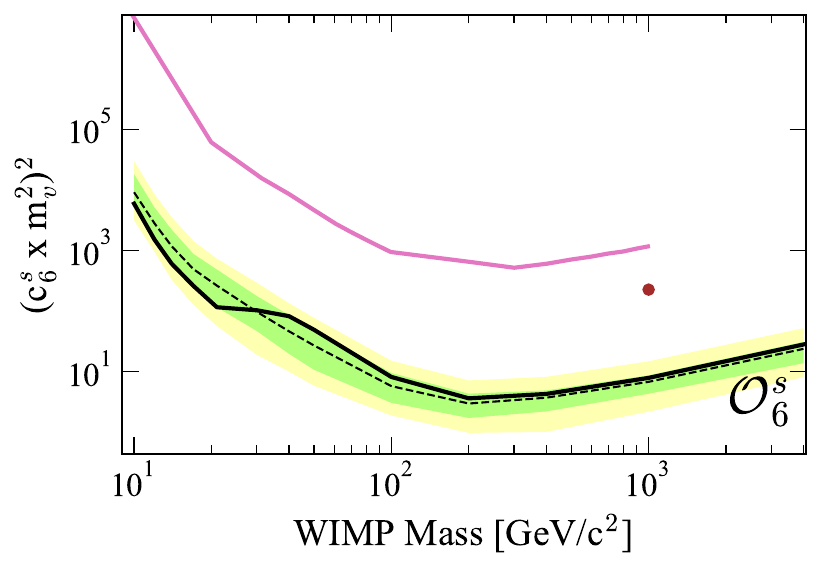}
    \includegraphics[width=0.5\columnwidth]{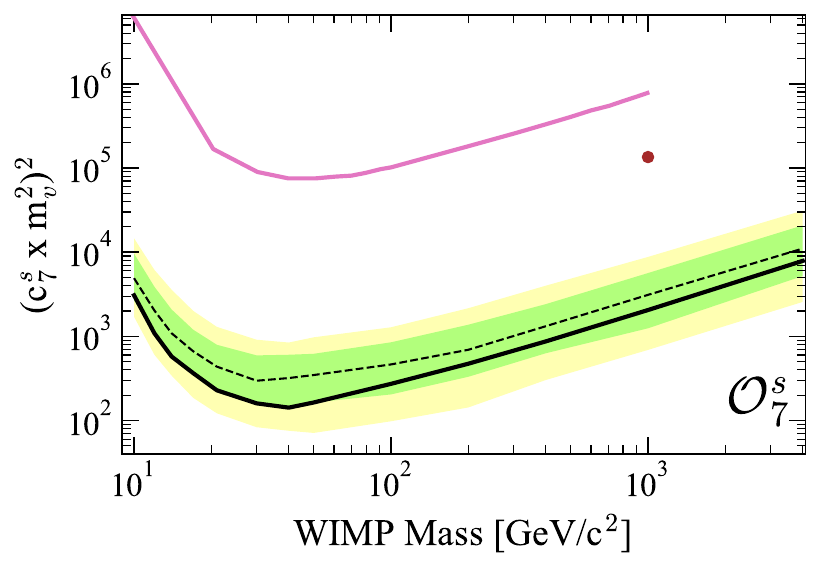}
    \includegraphics[width=0.5\columnwidth]{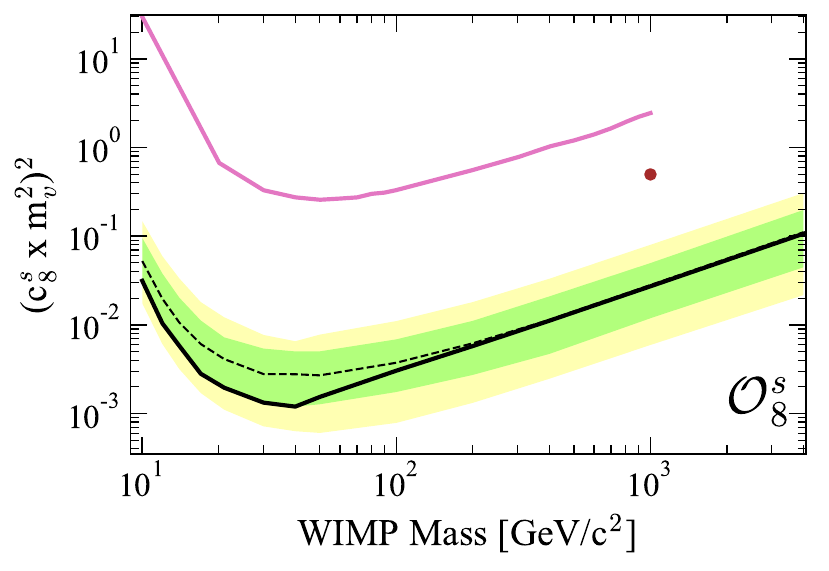}
    \includegraphics[width=0.5\columnwidth]{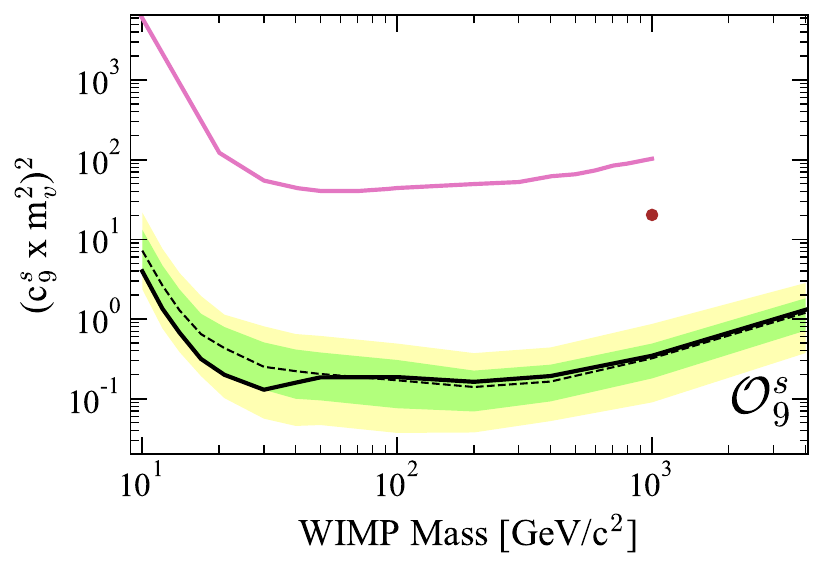}
    \includegraphics[width=0.5\columnwidth]{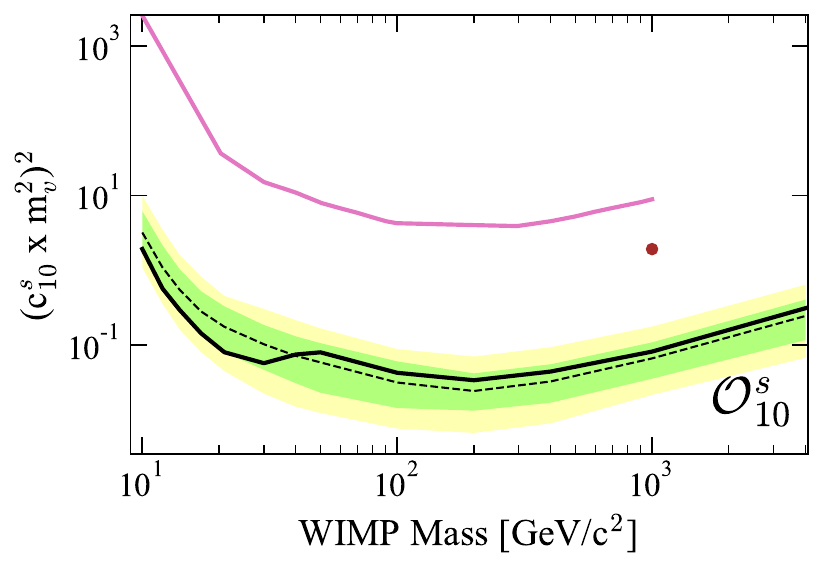}
    \includegraphics[width=0.5\columnwidth]{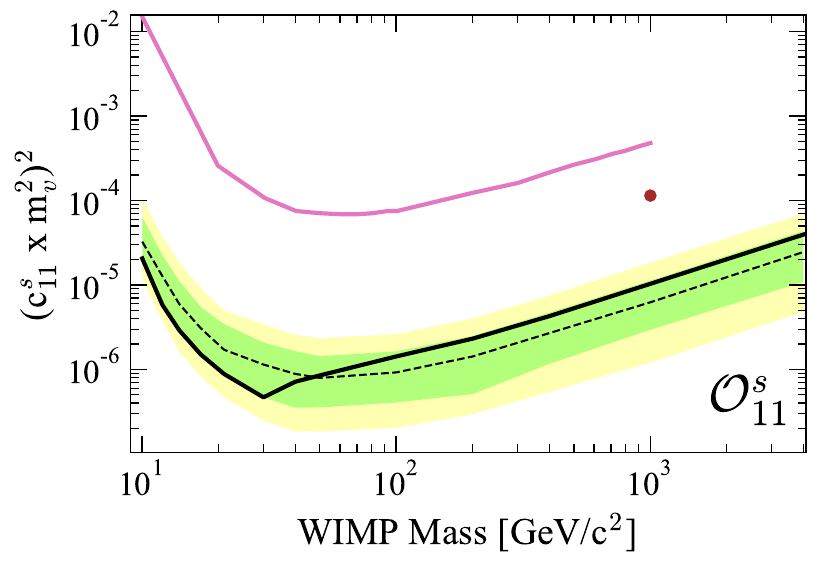}
    \includegraphics[width=0.5\columnwidth]{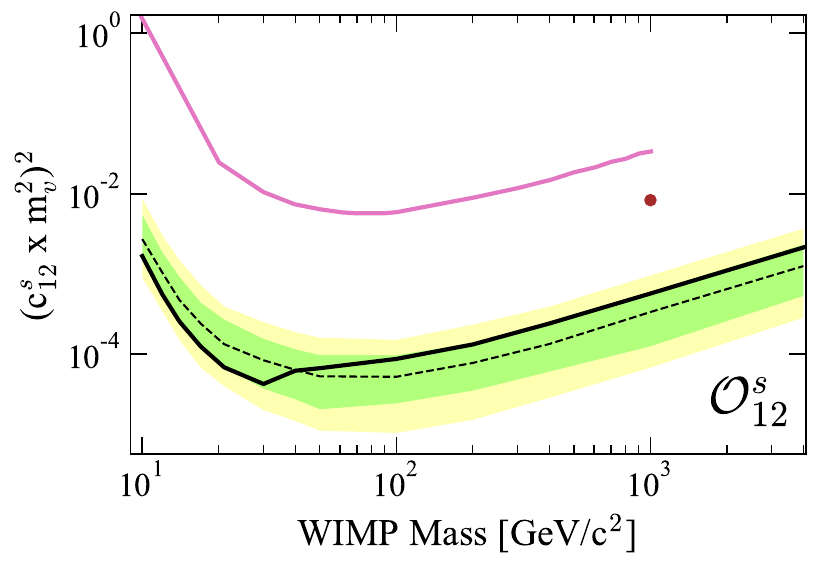}
    \includegraphics[width=0.5\columnwidth]{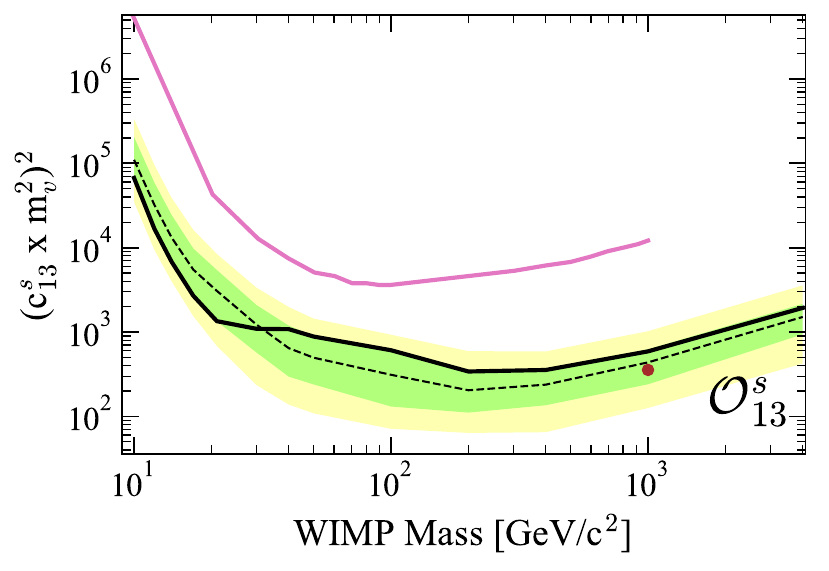}
    \includegraphics[width=0.5\columnwidth]{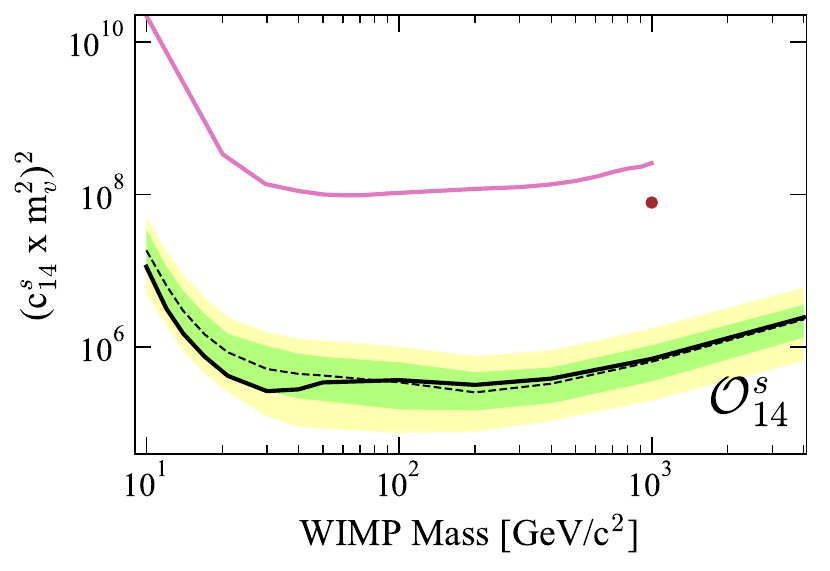}
    \includegraphics[width=0.5\columnwidth]{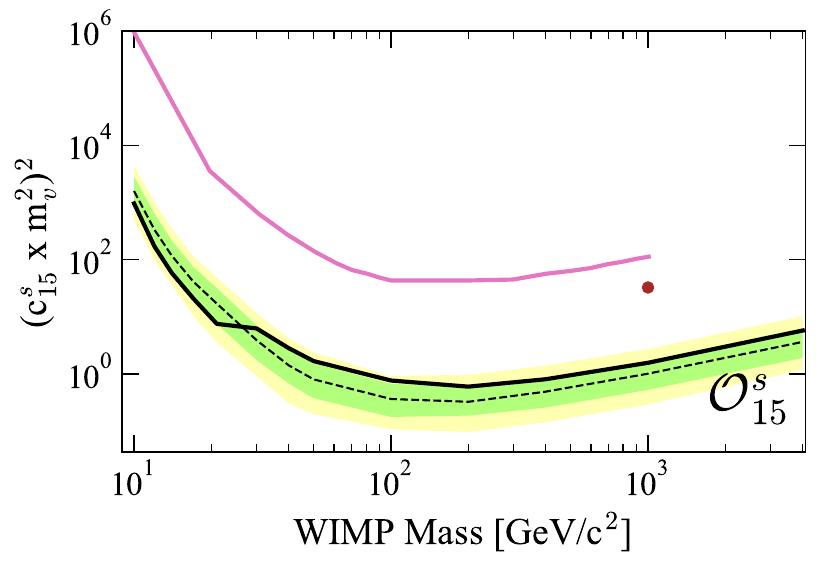}
    \caption{
    The 90\% confidence limit (black lines) of the dimensionless isoscalar WIMP-nucleon couplings for each of the fourteen NREFT elastic interactions.
    The black dotted lines show the medians of the sensitivity projection, and the green and yellow bands correspond to the 1$\sigma$ and 2$\sigma$ sensitivity bands, respectively.
    Also shown are the NREFT results from the XENON100 experiment (magenta) and the PandaX-II experiment (blue).
    The latter upper limits are cast from their starting points of $\mathcal{L}_5$ (reduces to $\mathcal{O}_1$) and $\mathcal{L}_{15}$ (reduces to $\mathcal{O}_4$).
    The LUX limits (brown points), are from their $\delta=0$~keV inelastic result.
    The LZ signal model uses nuclear density matrices that have been updated since the XENON100 and LUX analyses, leading to reduced rates for some operators such as $\mathcal{O}_{13}$, where the LUX result appears stronger than the LZ result. 
    }
  \label{fig:limits-elastic-s}
\end{figure*}
\begin{figure*}
    \includegraphics[width=0.5\columnwidth]{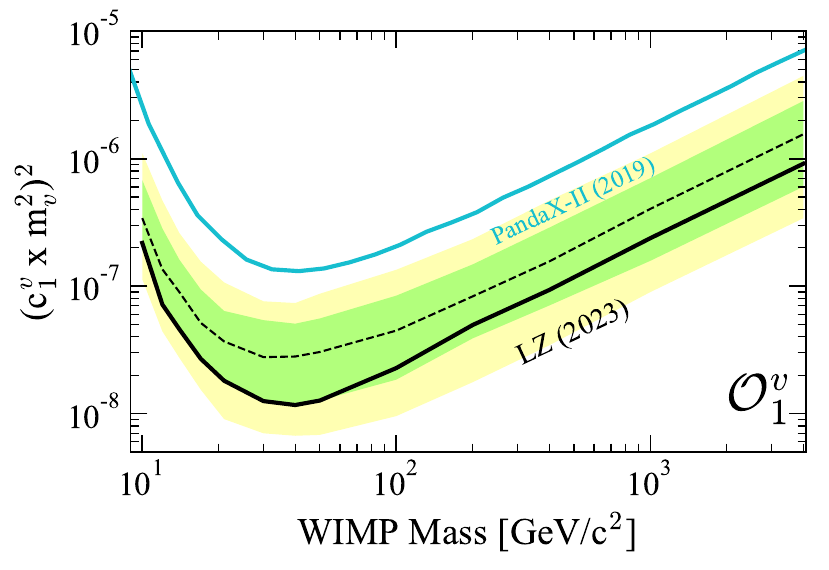}
    \includegraphics[width=0.5\columnwidth]{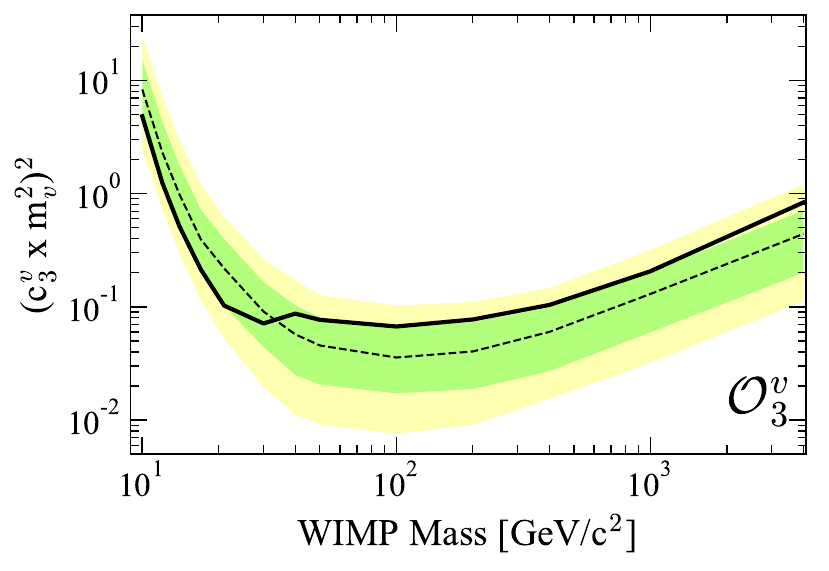}
    \includegraphics[width=0.5\columnwidth]{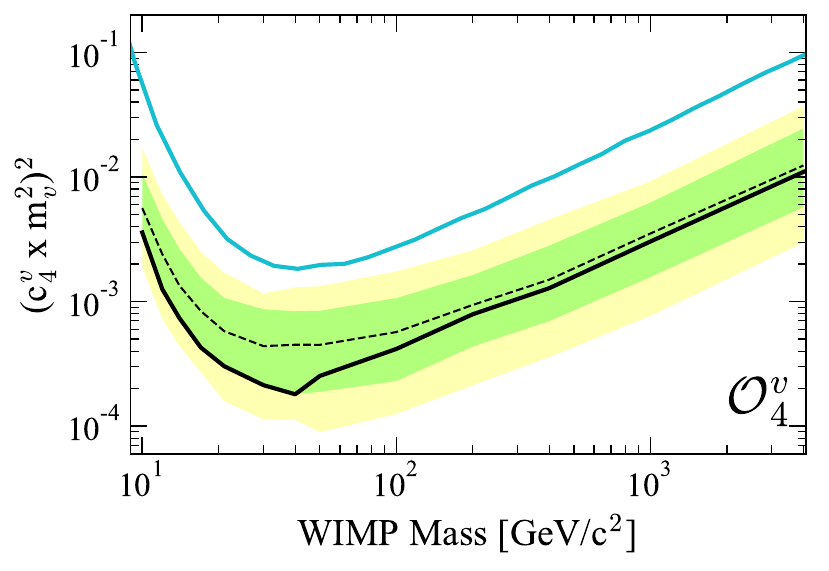}
    \includegraphics[width=0.5\columnwidth]{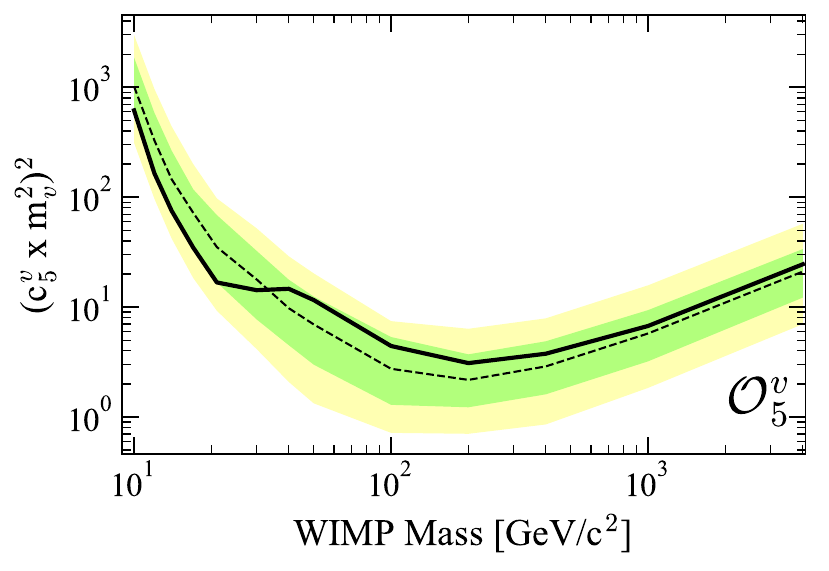}
    \includegraphics[width=0.5\columnwidth]{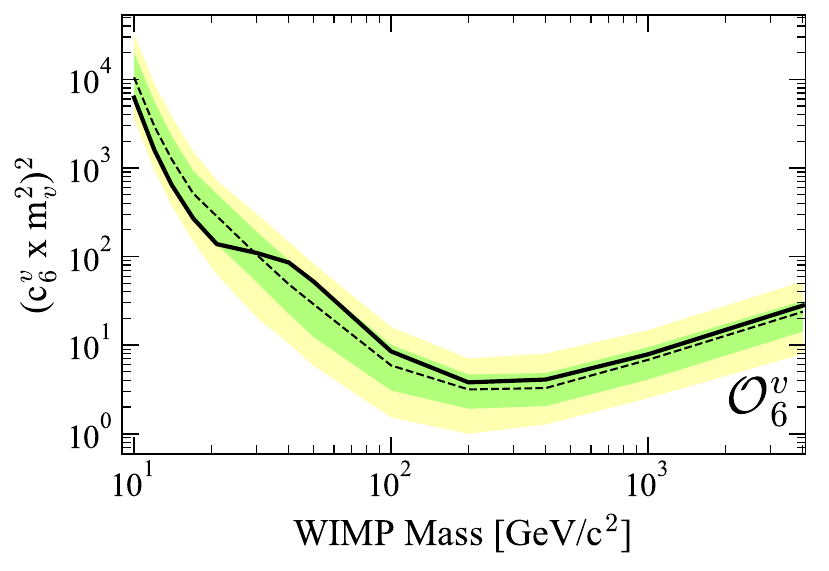}
    \includegraphics[width=0.5\columnwidth]{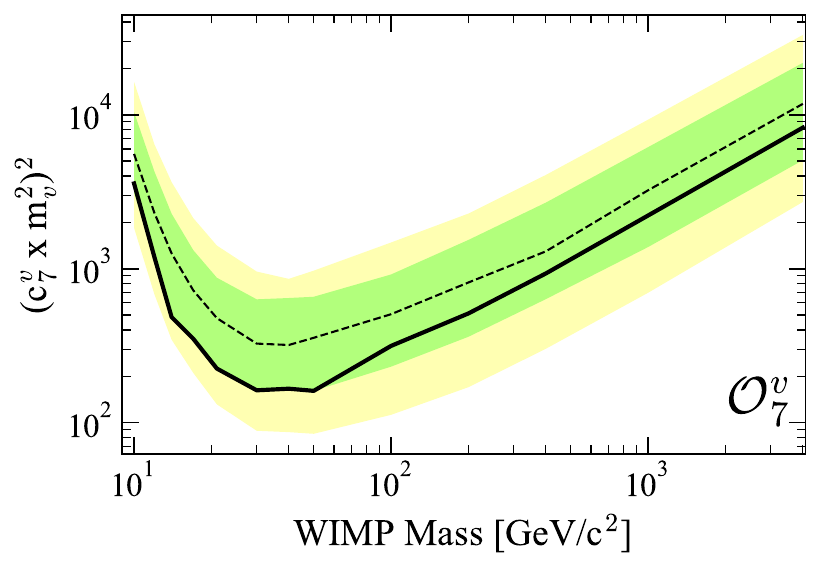}
    \includegraphics[width=0.5\columnwidth]{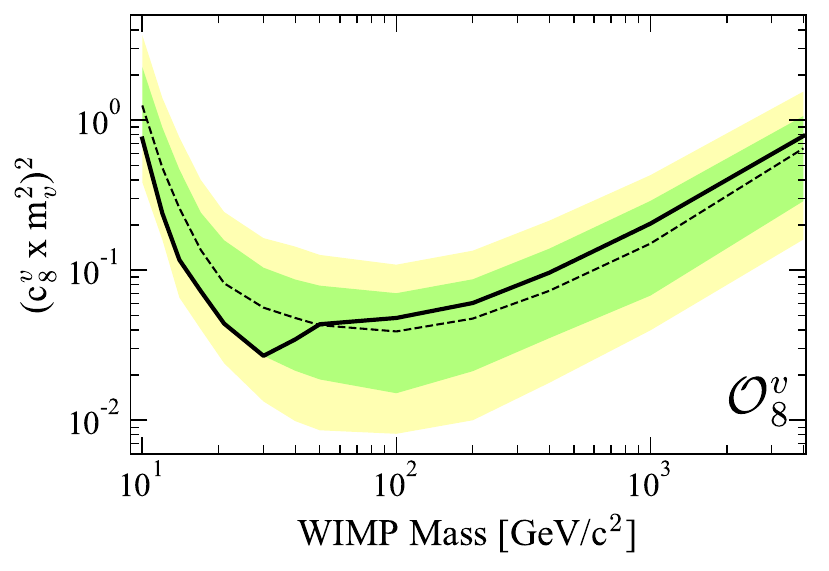}
    \includegraphics[width=0.5\columnwidth]{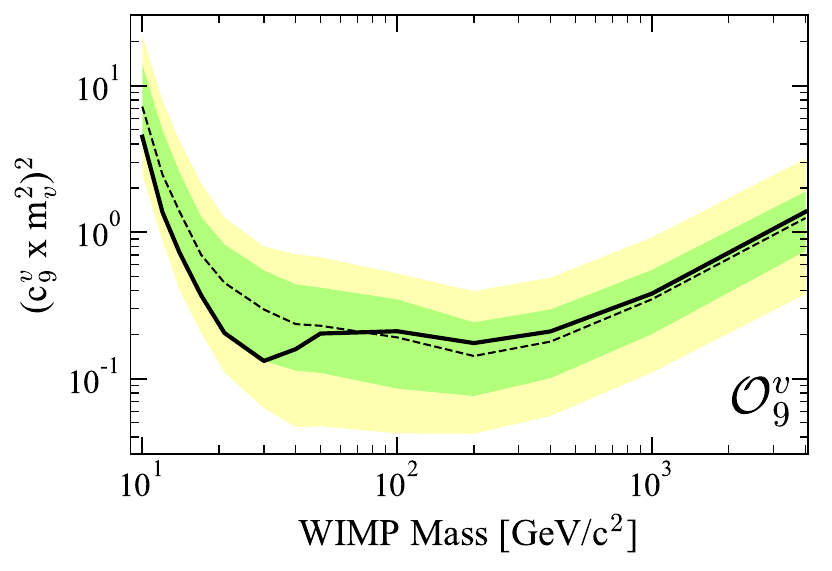}
    \includegraphics[width=0.5\columnwidth]{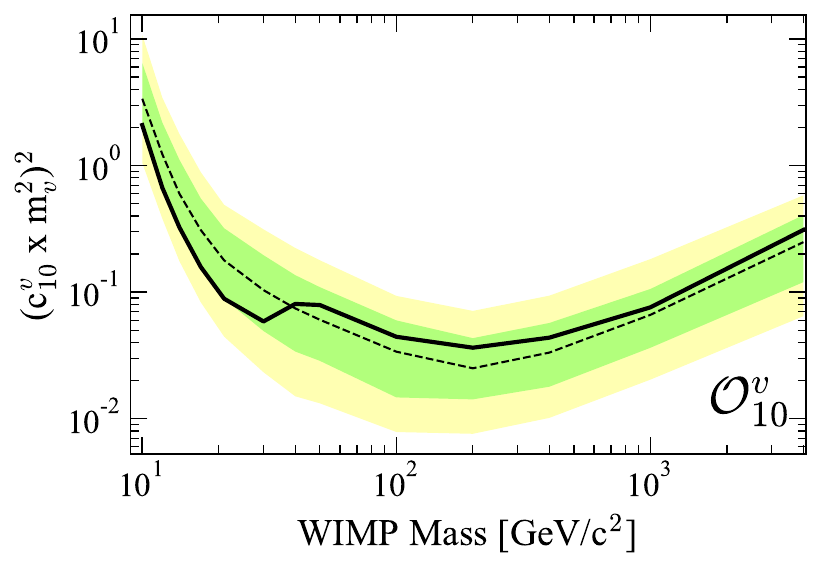}
    \includegraphics[width=0.5\columnwidth]{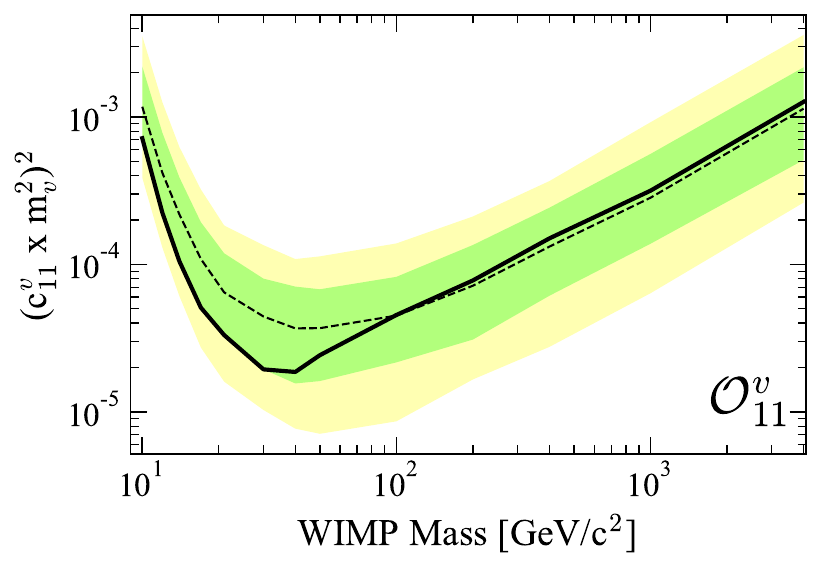}
    \includegraphics[width=0.5\columnwidth]{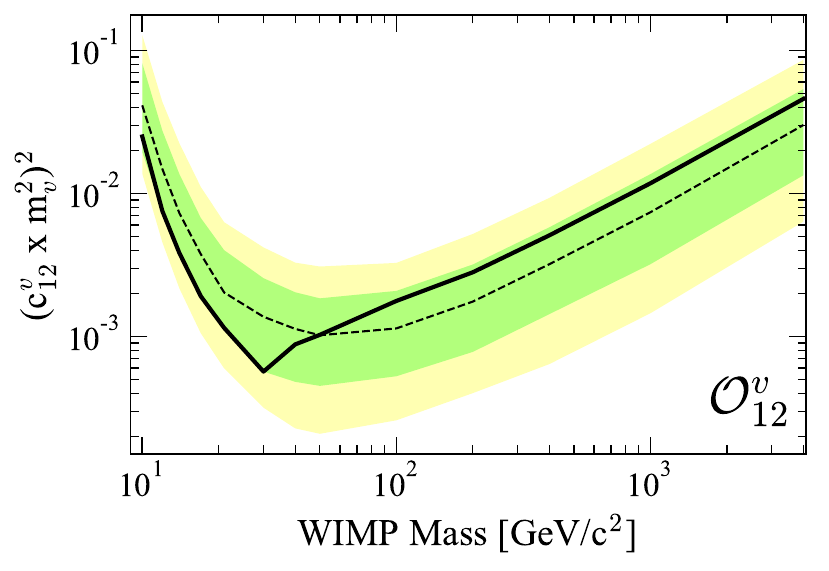}
    \includegraphics[width=0.5\columnwidth]{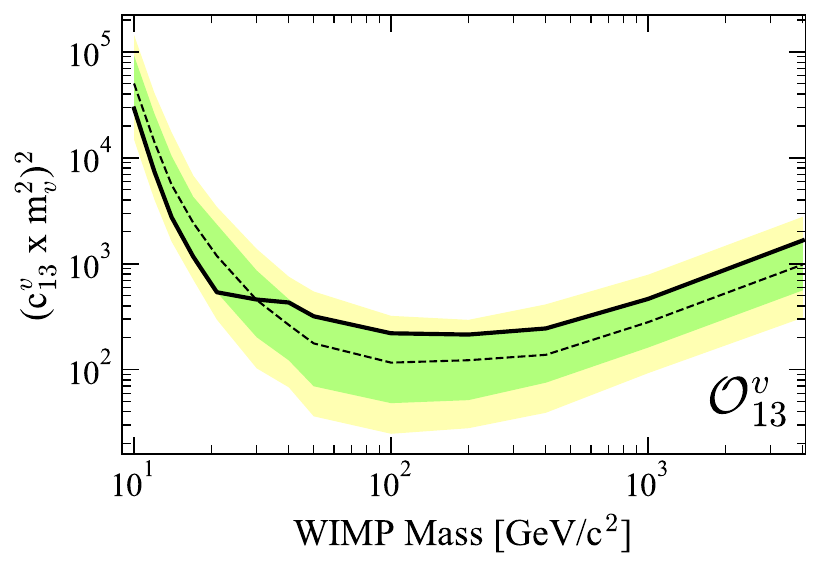}
    \includegraphics[width=0.5\columnwidth]{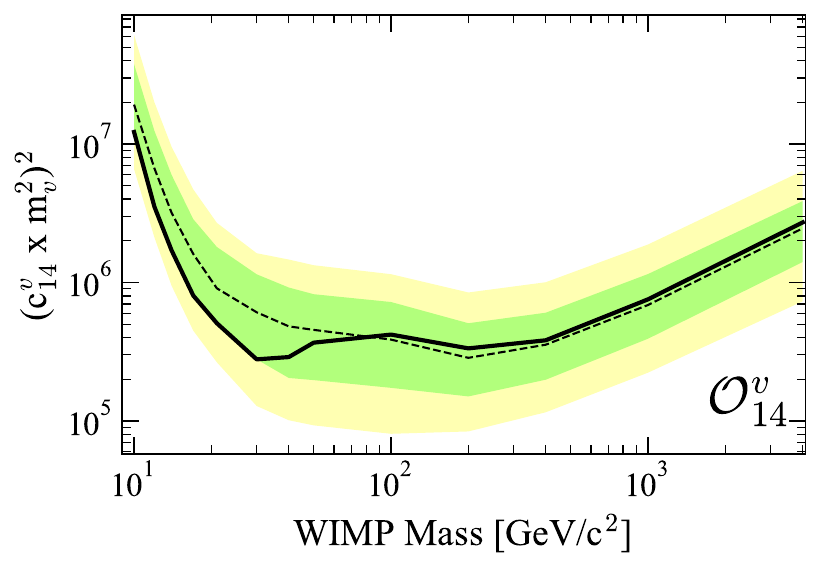}
    \includegraphics[width=0.5\columnwidth]{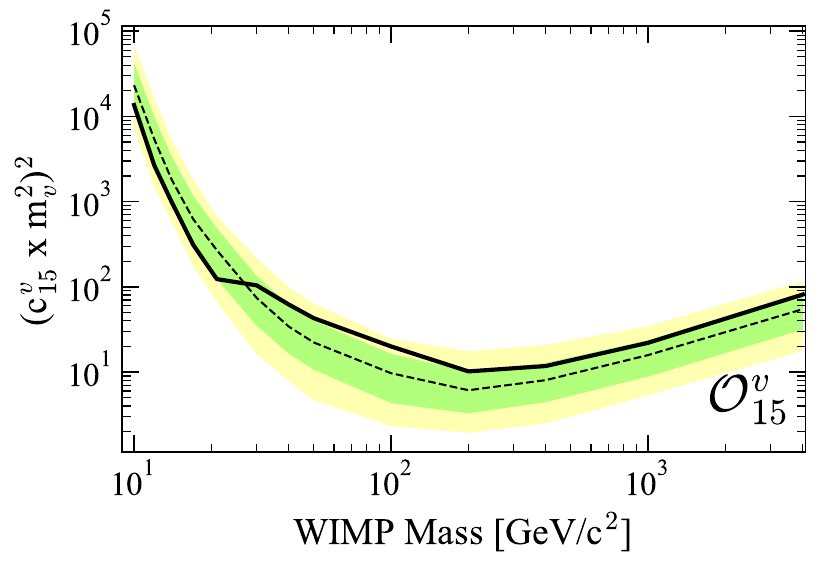}
    \caption{The 90\% confidence limit (black lines) of the dimensionless isovector WIMP-nucleon couplings for each of the fourteen NREFT elastic interactions.
    The black dotted lines show the medians of the sensitivity projection, and the green and yellow bands correspond to the 1$\sigma$ and 2$\sigma$ sensitivity bands, respectively.
    Also shown are the SI and SD results of the PandaX-II experiment (blue).
    The latter upper limits are cast from their $\mathcal{L}_5$ (reduces to $\mathcal{O}_1$) and their $\mathcal{L}_{15}$ (reduces to $\mathcal{O}_4$)}
  \label{fig:limits-elastic-v}
\end{figure*}

The upper limits for the coupling strengths of DM scattering elastically via the operators $\mathcal{O}_{1,3-15}$ are shown in \autoref{fig:limits-elastic-s} and \autoref{fig:limits-elastic-v} for isoscalar and isovector interactions, respectively. 
A power constraint is applied on all operators (in the 17--30~GeV/c$^2$ mass range) to restrict the upper limit falling 1$\sigma$ below the median expectation due to background fluctuations. 
The constrained limit corresponded to a critical alternate hypothesis power of $\pi_\text{crit} = 0.16$ \cite{LZ:SR1WS_2022, DM_parameters:BAXTER2021_Conventions, Cowan:2011_power_constraints}.
Observed upper limits are compatible with background-only expectations to within $1\sigma$ for the majority of the operator-mass combinations.
A few operator-mass combinations, e.g. $\mathcal{O}_{3, 13, 15}$ above 30~GeV/c$^2$, are found to have limits weaker than the median expectation.
These discrepancies are within $2\sigma$ and are generally caused by ER background leakage into the NR region occupied by these momentum-suppressed operators with highly peaking spectra (see \autoref{fig:recoils}).
The deviation from the ER band of the two most egregious outlier events is consistent with unresolved multiple scatter from detector-based ER decays.
\par
Consistency with the first LZ result (Ref.~\cite{LZ:SR1WS_2022}) is established with $\mathcal{O}_1$, the SI operator that couples solely to the total number of nucleons.
Unaffected by the $\vec{v}^\perp$ and $\vec{q}$ degrees of freedom, $\mathcal{O}_1$ and $\mathcal{O}_4$ yield some of the most stringent constraints on the couplings of nearly all the operators.
The most stringent limit is set by $\mathcal{O}_1$ and the limit on $\mathcal{O}_4$ is only exceeded by those of $\mathcal{O}_{11}$ and $\mathcal{O}_{12}$. 
The nuclear response of $\mathcal{O}_{11}$ is similar to that of the SI operator $\mathcal{O}_1$ at higher energies, however momentum-dependence causes a suppressed rate at low energies. 
$\mathcal{O}_{12}$ is an example of an operator for which the rate is enhanced by positive parity nucleon velocity contributions that are summed over the composite nucleus~\cite{Anand:MathematicaEFT}. 
While, the exclusion curves for most operators have minima at WIMP masses of 30--50~GeV/c$^2$, interactions that are suppressed by two powers of $\vec{q}$ such as $\mathcal{O}_6$ and $\mathcal{O}_{15}$ attained minima in their coupling strengths at WIMP masses of 200--300~GeV/c$^2$; these are the operators that benefit the most from an extended energy window.
\par
Results of the XENON100~\cite{Xenon100:EFT_2017}, LUX~\cite{LUX:EFTR4_2021}, and PandaX~\cite{PandaX2:SD_EFT_2019} isoscalar analyses are also shown in \autoref{fig:limits-elastic-s} for comparison. 
The LUX constraints are only available for a WIMP mass of 1~TeV/c$^2$ since they originate from an inelastic DM search from which the data for zero mass splitting were used.
All results are normalized to the dimensionless form $(c_i^N \times m_v^2)^2$, and the differing normalization conventions among experiments are presented in \cref{ap:1}. 
Several other results are omitted in the comparison, primarily due to their choice of presenting limits in the proton and neutron bases instead of the isoscalar and isovector bases.
Previous searches for NREFT interactions also used one-body nuclear density matrices for Xe that have since been updated \cite{Haxton_OneBody}. 
This analysis uses the updated matrices with the effect of altered event rates for some operators, notably a decrease in the event rate for $\mathcal{O}_{13}$, leading to an upper limit weaker than previous results in some cases. 

\subsection{Inelastic}\label{subsec:inelastic}
\begin{figure*}
    \includegraphics[width=0.5\columnwidth]{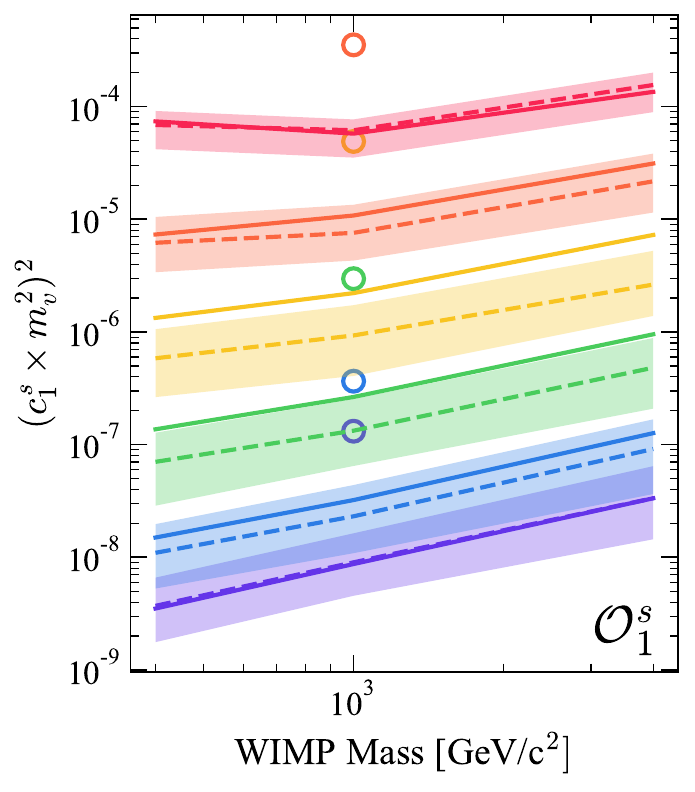}
    \includegraphics[width=0.5\columnwidth]{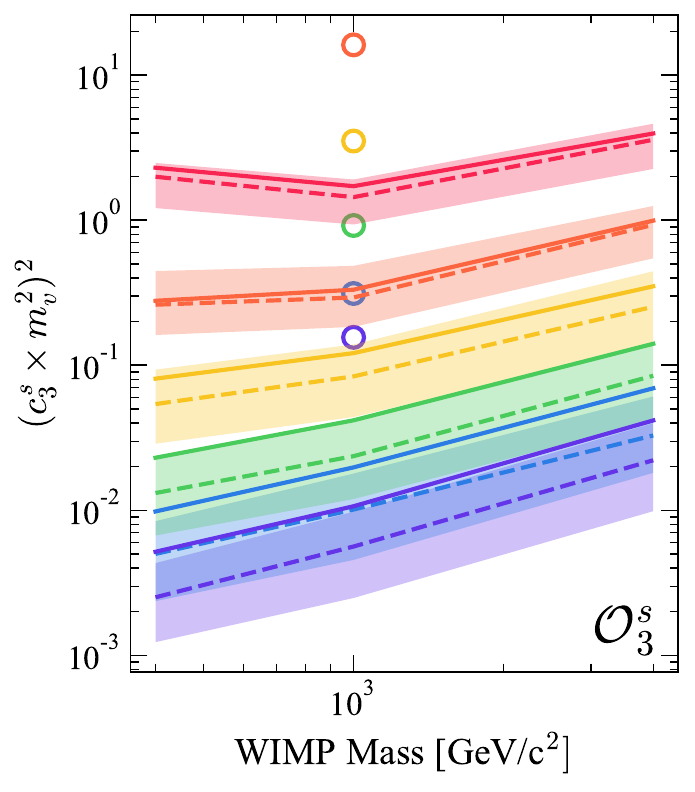}
    \includegraphics[width=0.5\columnwidth]{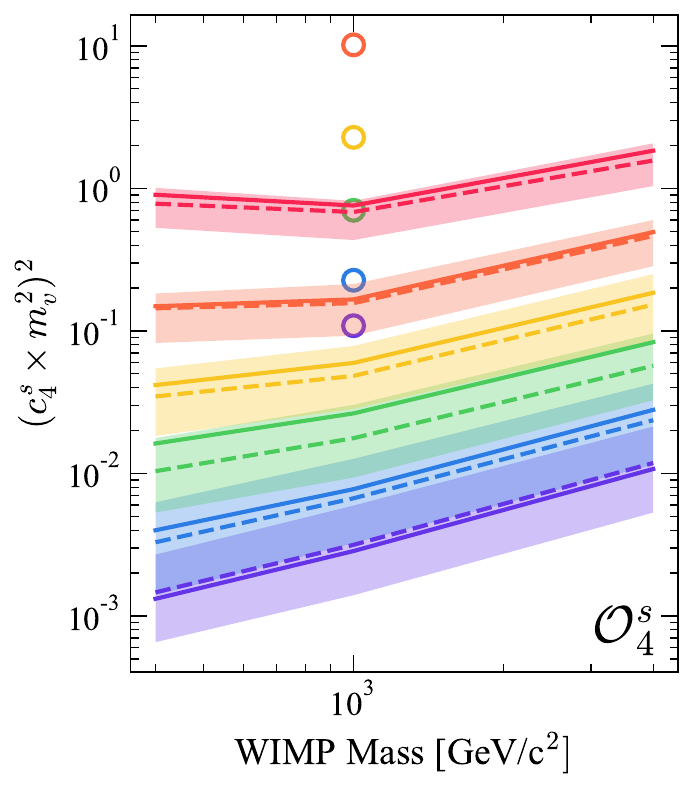}
    \includegraphics[width=0.5\columnwidth]{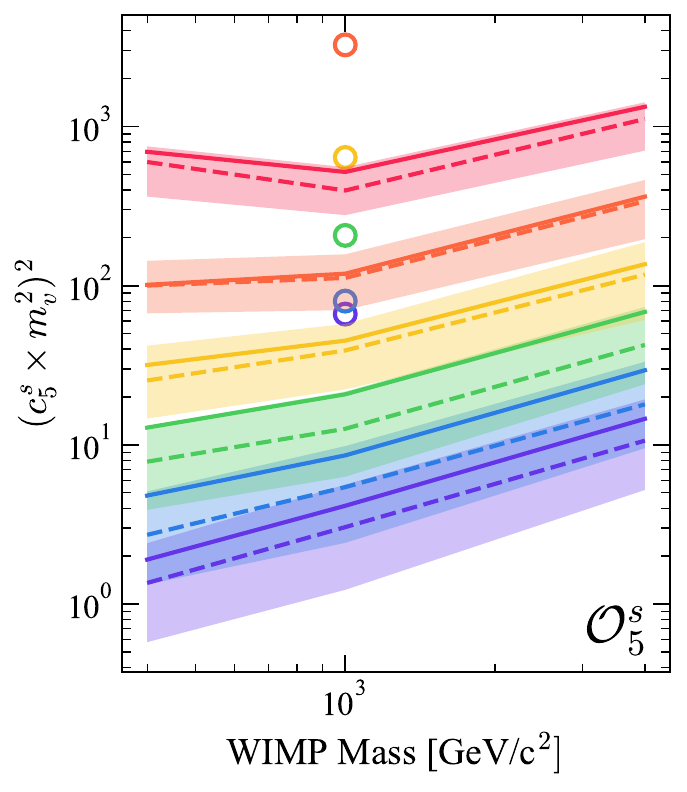}
    \includegraphics[width=0.5\columnwidth]{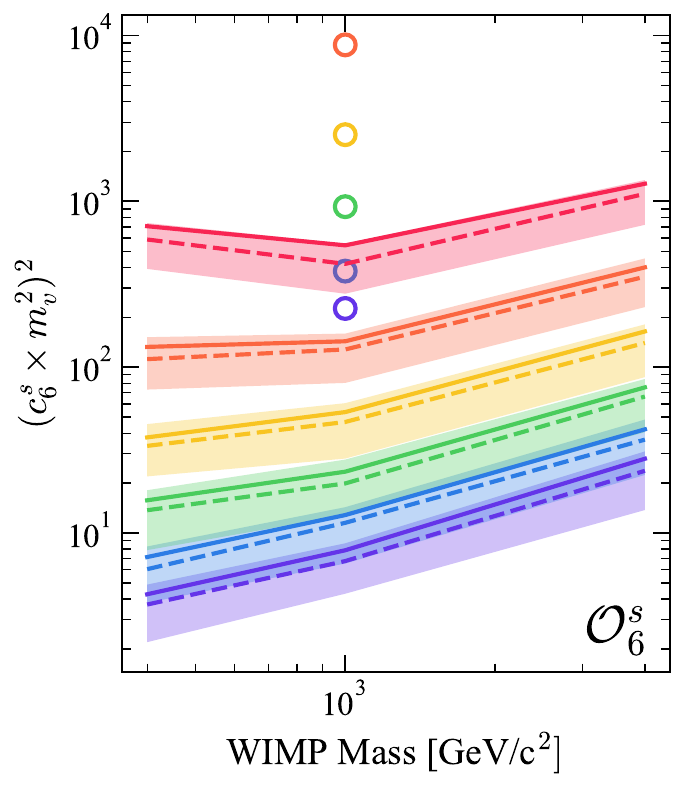}
    \includegraphics[width=0.5\columnwidth]{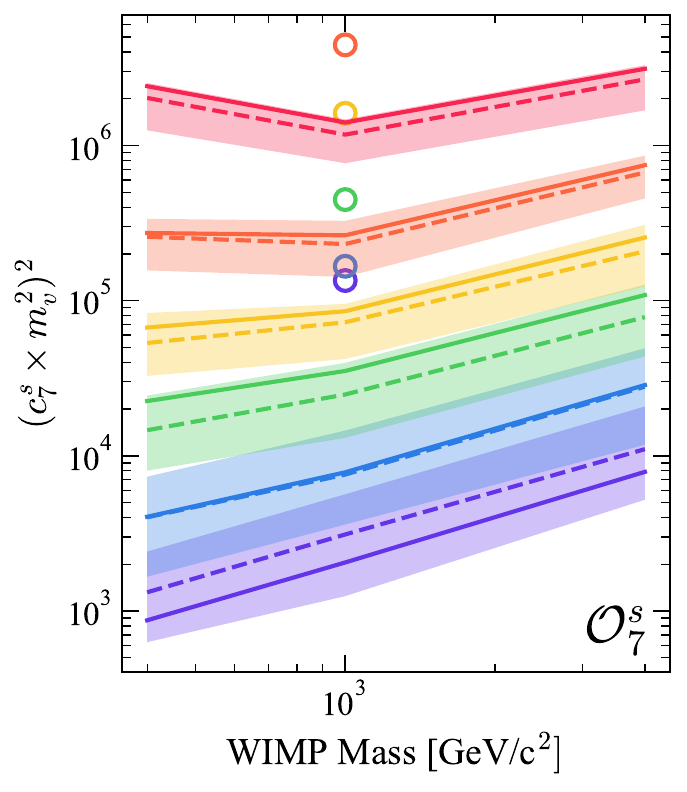}
    \includegraphics[width=0.5\columnwidth]{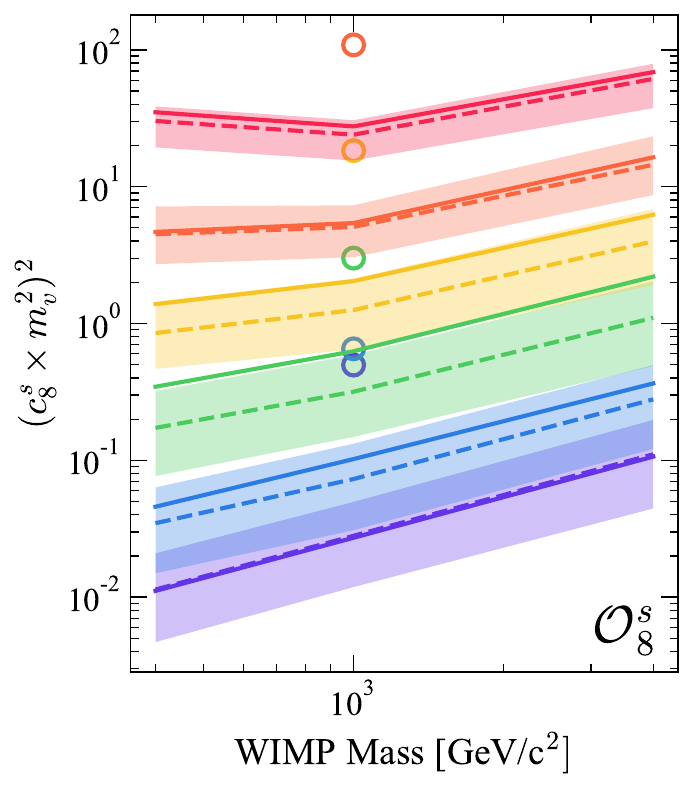}
    \includegraphics[width=0.5\columnwidth]{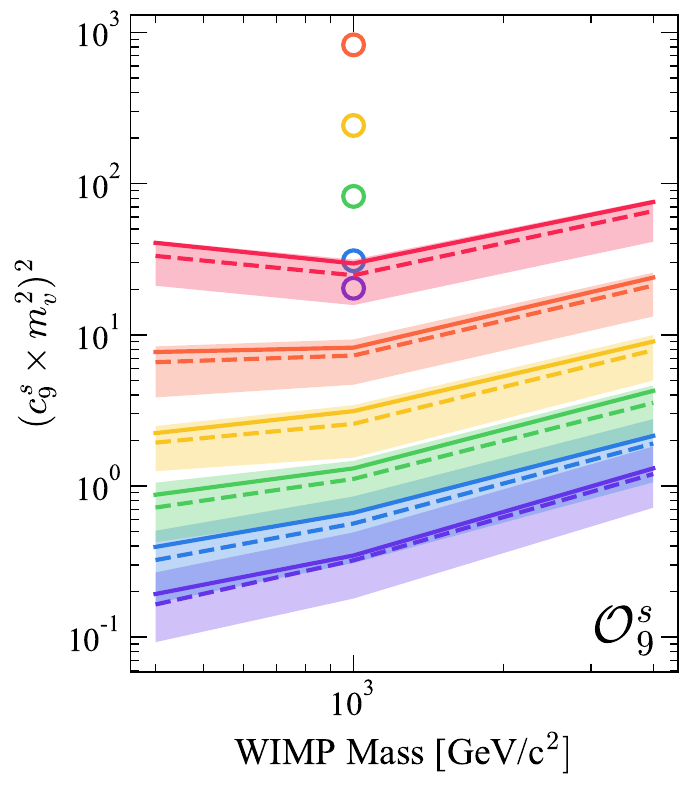}
    \includegraphics[width=0.5\columnwidth]{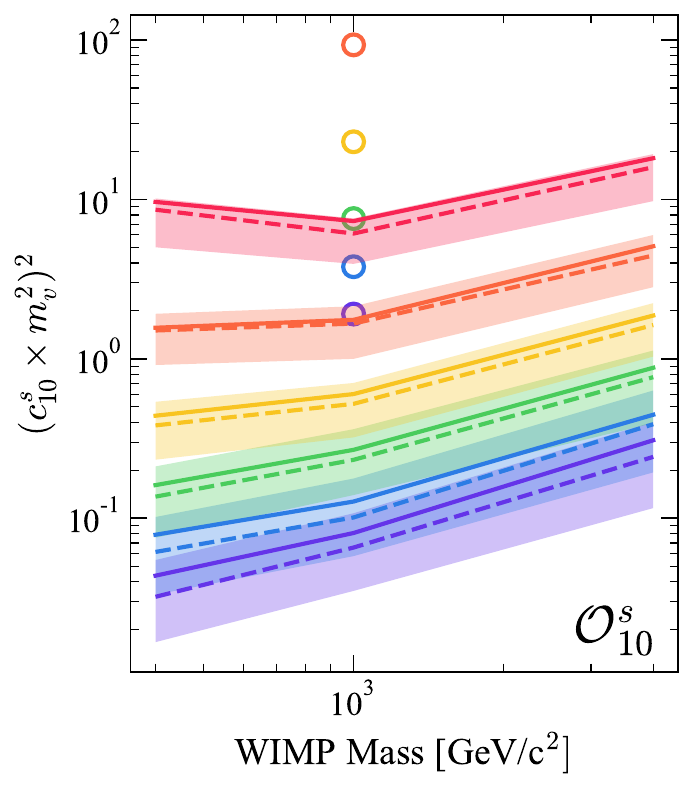}
    \includegraphics[width=0.5\columnwidth]{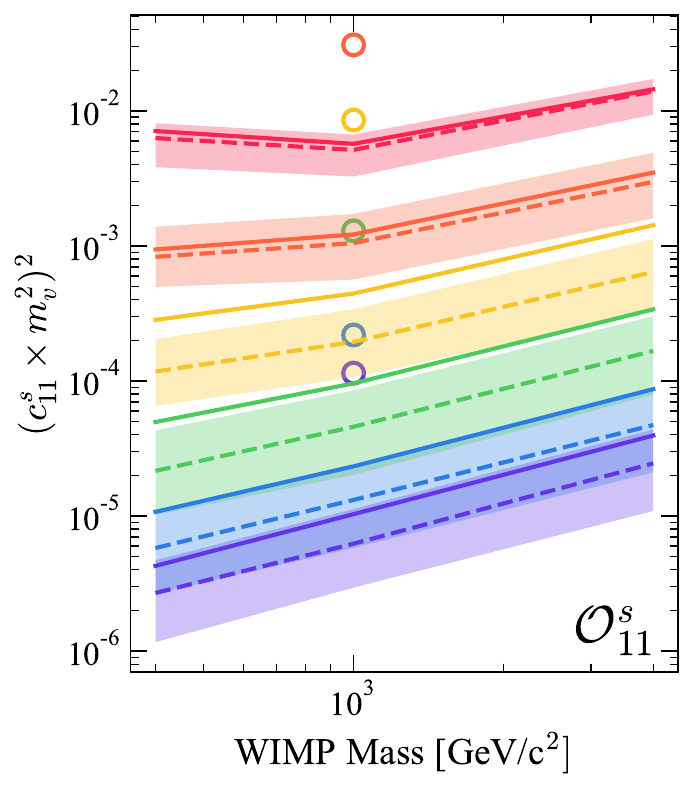}
    \includegraphics[width=0.5\columnwidth]{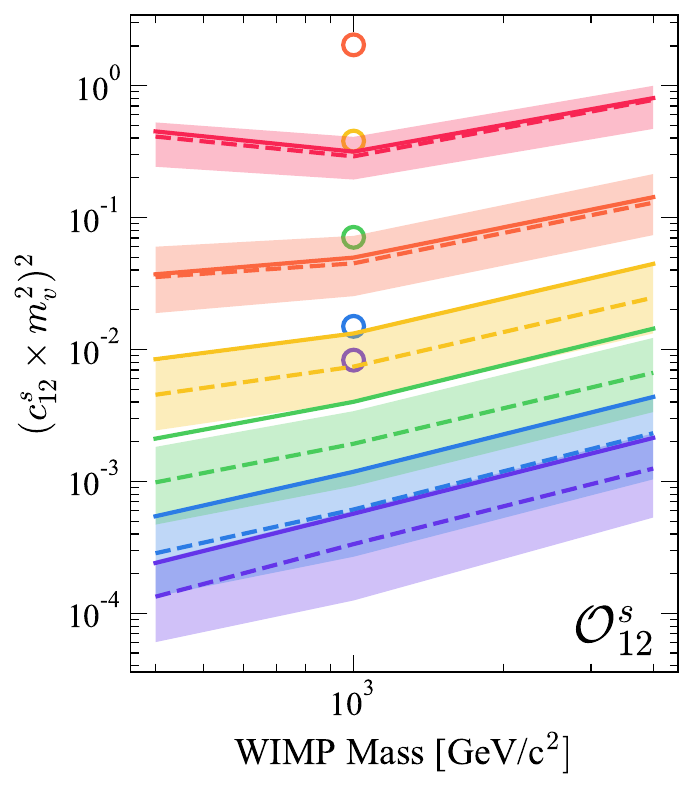}
    \includegraphics[width=0.5\columnwidth]{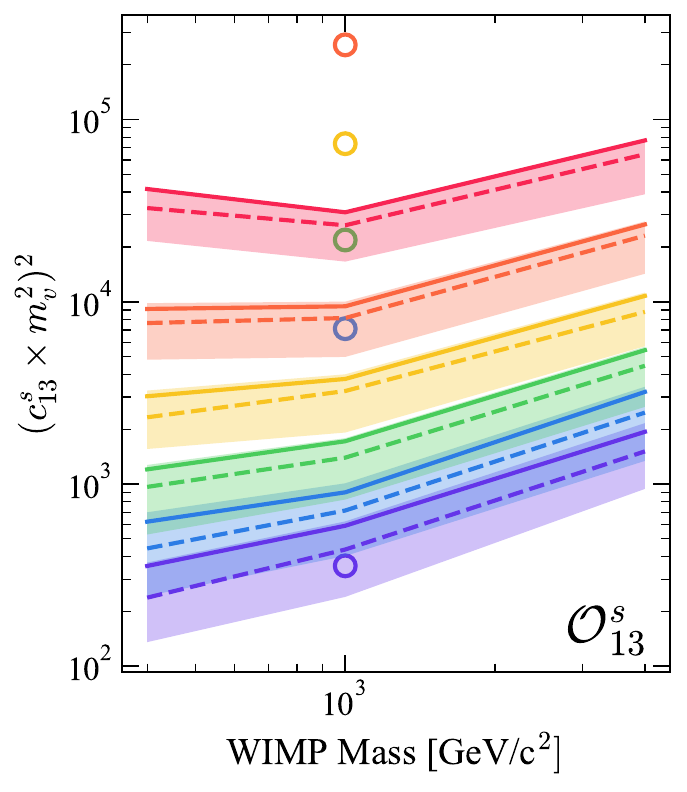}
    \includegraphics[width=0.5\columnwidth]{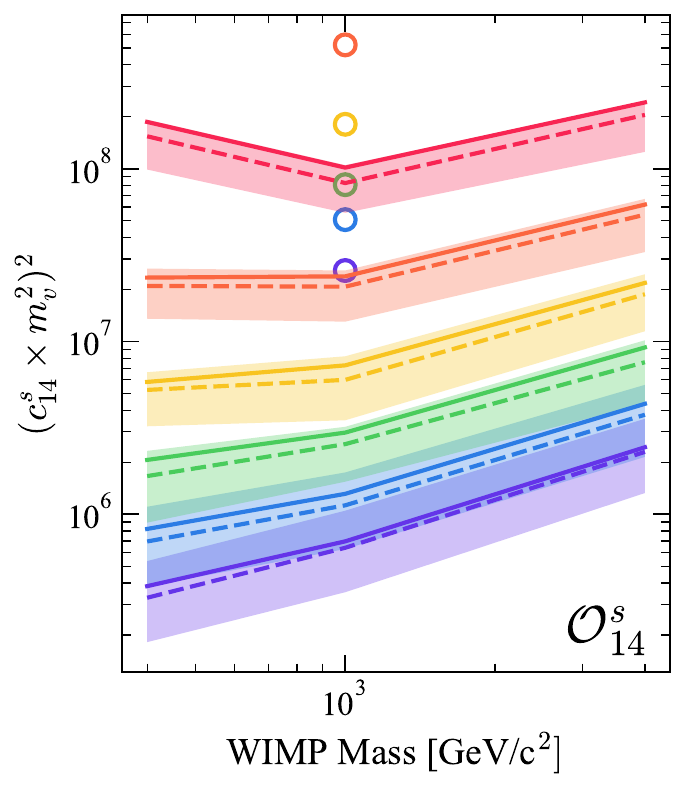}
    \includegraphics[width=0.5\columnwidth]{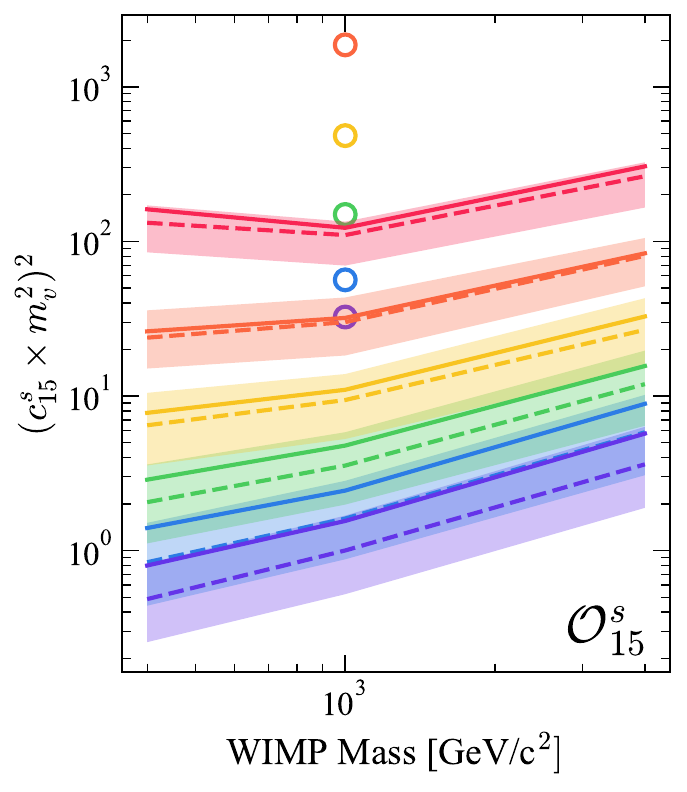}
    \includegraphics[width=0.25\columnwidth,trim=-5cm -20cm 0 3cm]{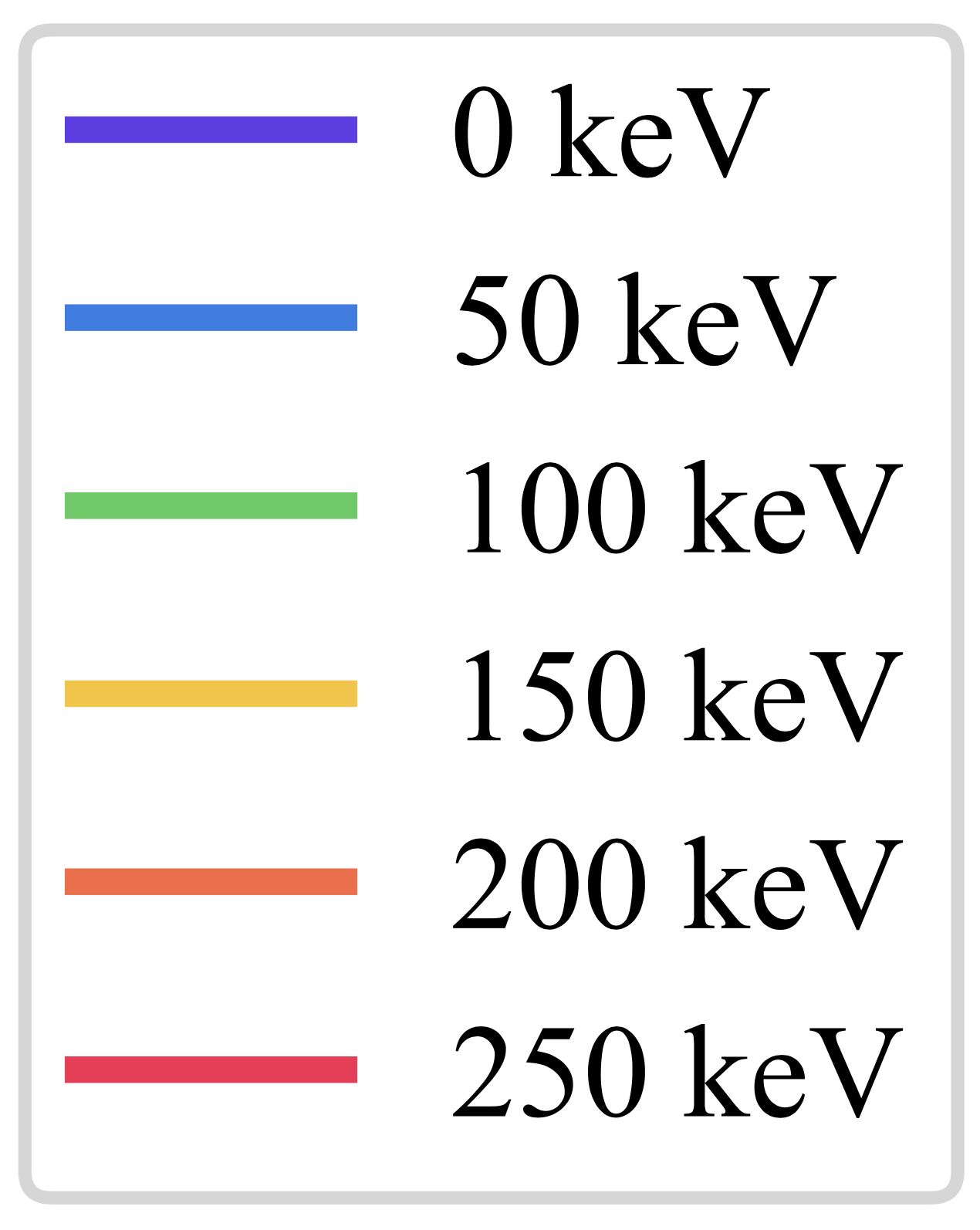}
    \caption{ The 90\% confidence limit (solid lines) of the dimensionless isoscalar WIMP-nucleon couplings for each of the fourteen NREFT interactions.
    The dotted lines show the medians of the sensitivity projection and the shaded bands correspond to the 1$\sigma$ sensitivity band.
    The upper limit is evaluated for WIMP masses of 400~GeV/c$^2$, 1000~GeV/c$^2$, and 4000~GeV/c$^2$, for values of the mass splitting $\delta$ of 0~keV (purple), 50~keV (blue), 100~keV (green), 150~keV (yellow), 200~keV (orange), and 250~keV (red).
    Circular data points represent the 1000~GeV/c$^2$ inelastic limits from the LUX WS2014-16 NREFT search~\cite{LUX:EFTR4_2021}.}
    \label{fig:limits-inelastic-s}
\end{figure*}

\begin{figure*}
    \includegraphics[width=0.5\columnwidth]{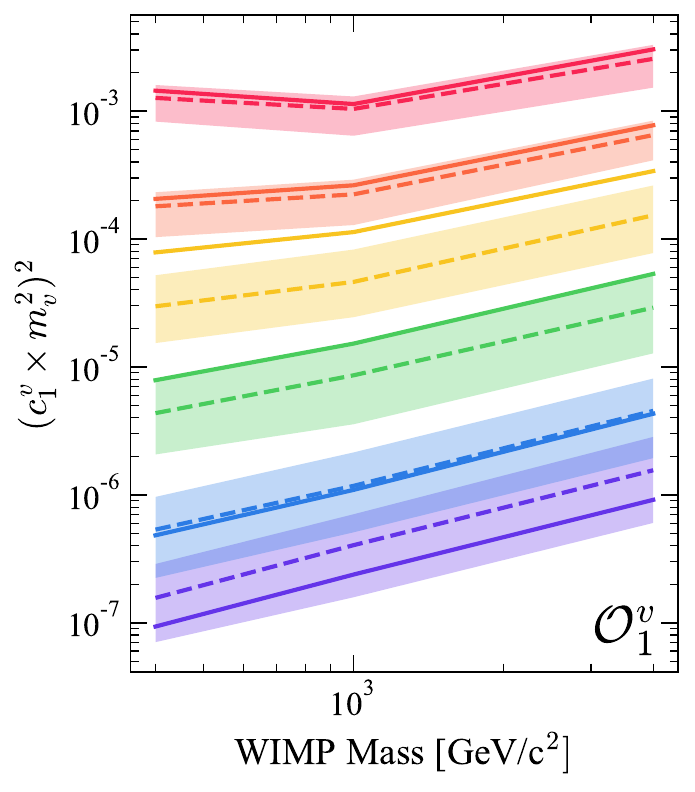}
    \includegraphics[width=0.5\columnwidth]{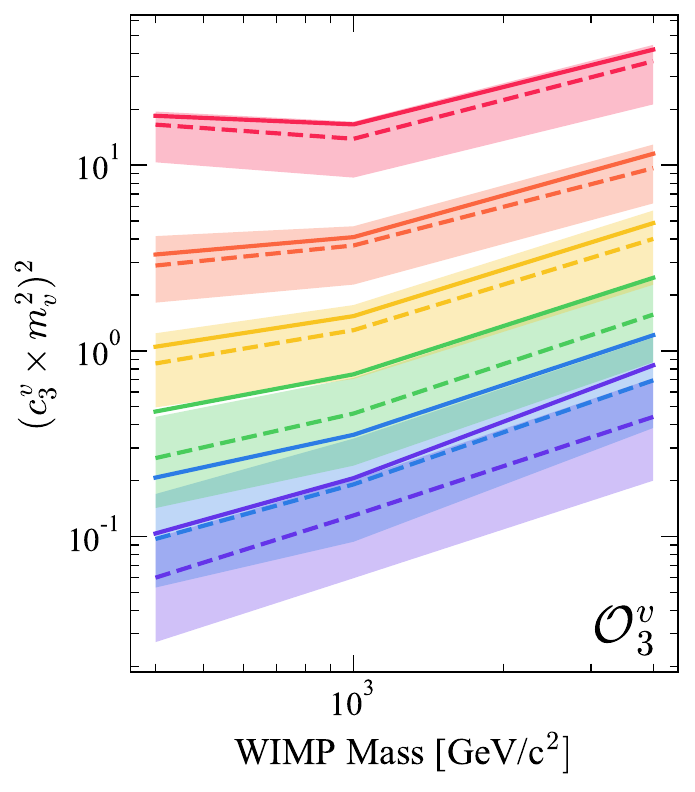}
    \includegraphics[width=0.5\columnwidth]{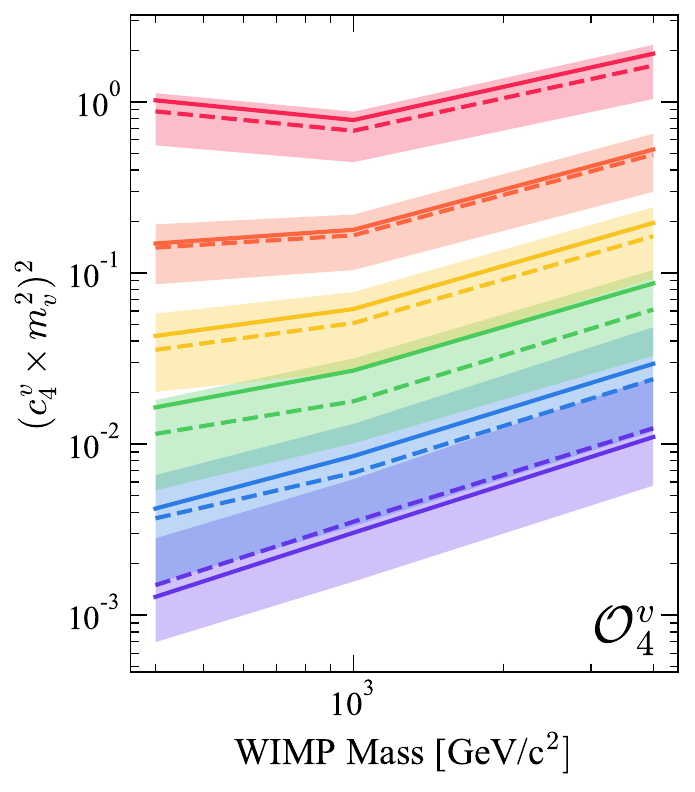}
    \includegraphics[width=0.5\columnwidth]{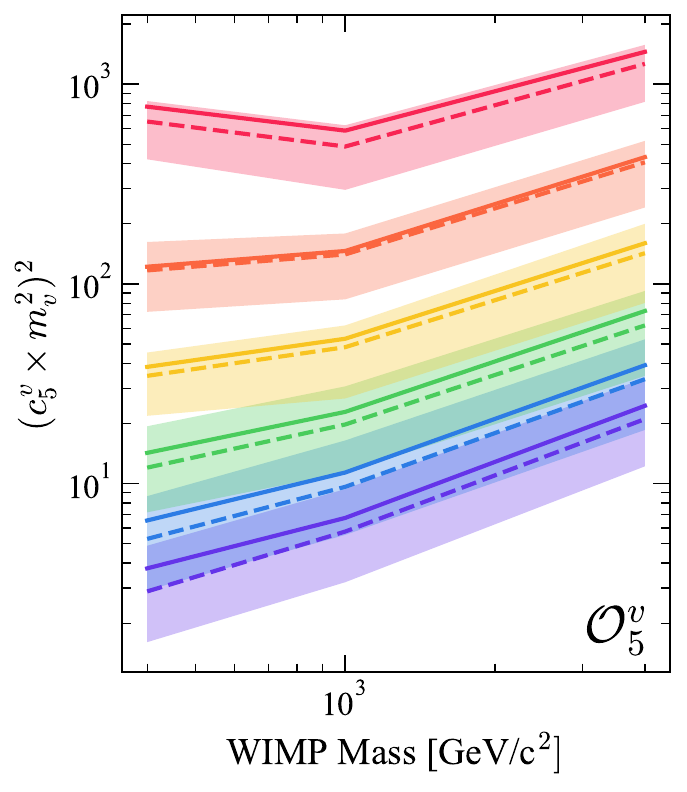}
    \includegraphics[width=0.5\columnwidth]{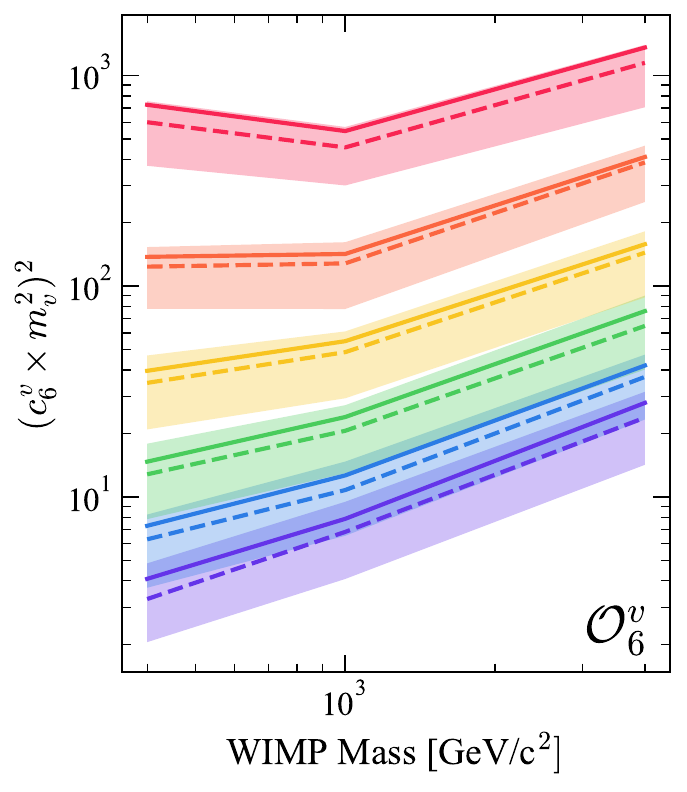}
    \includegraphics[width=0.5\columnwidth]{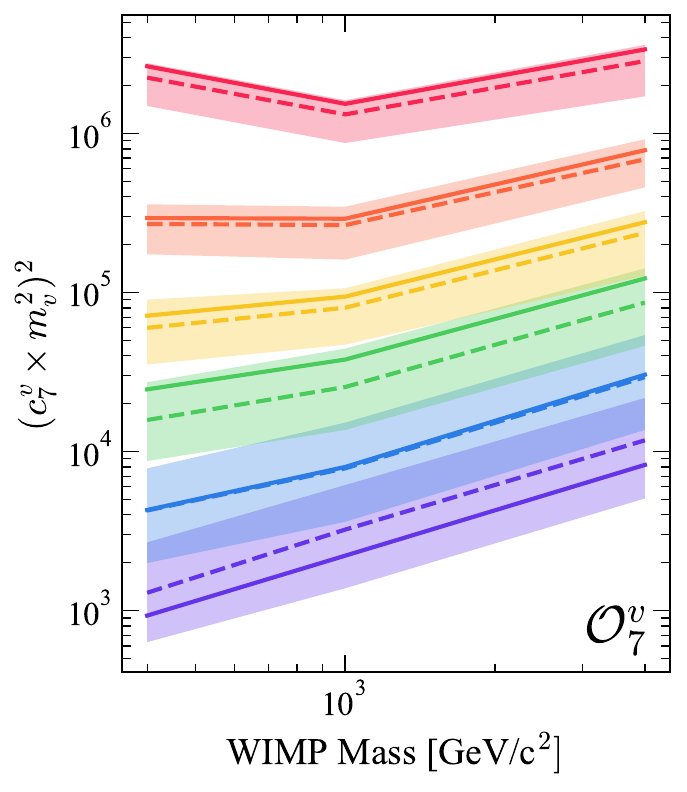}
    \includegraphics[width=0.5\columnwidth]{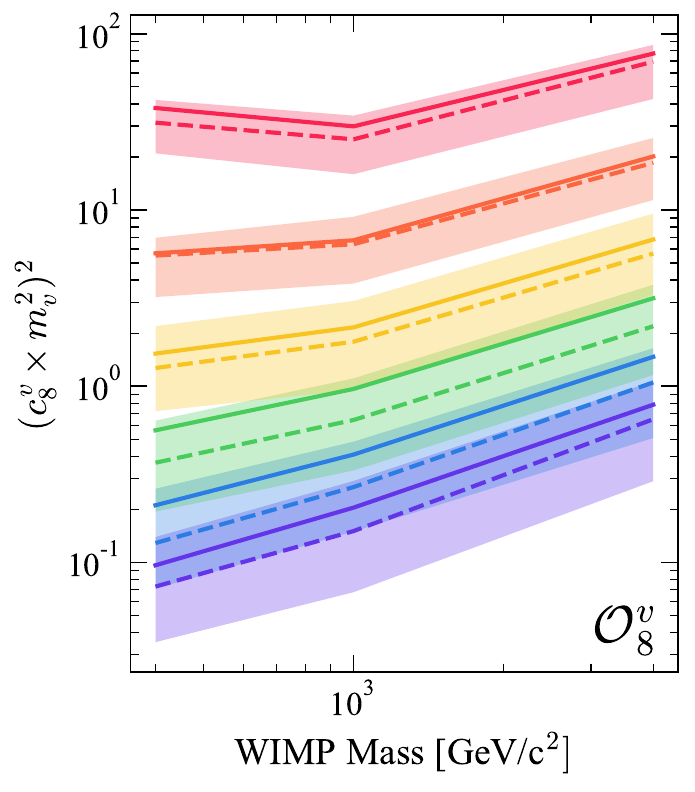}
    \includegraphics[width=0.5\columnwidth]{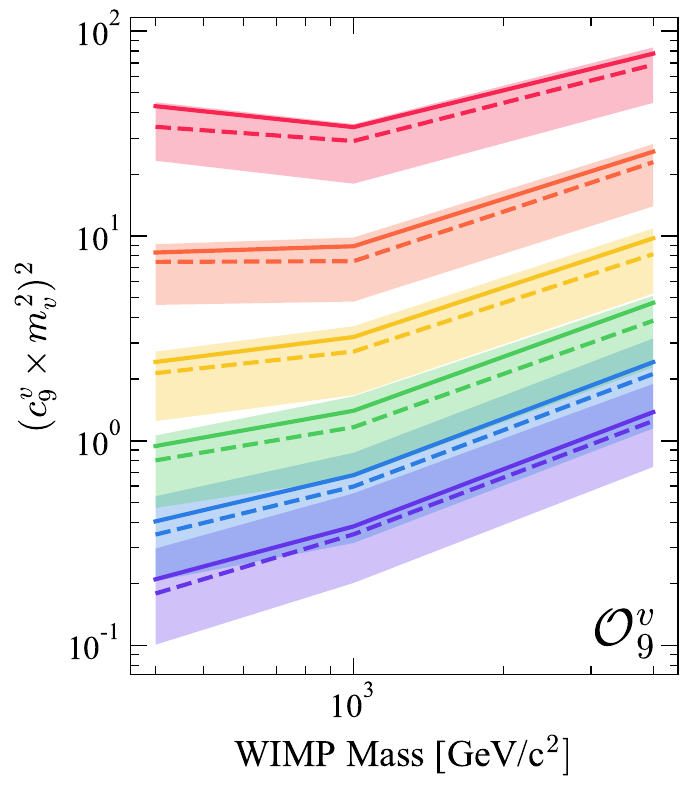}
    \includegraphics[width=0.5\columnwidth]{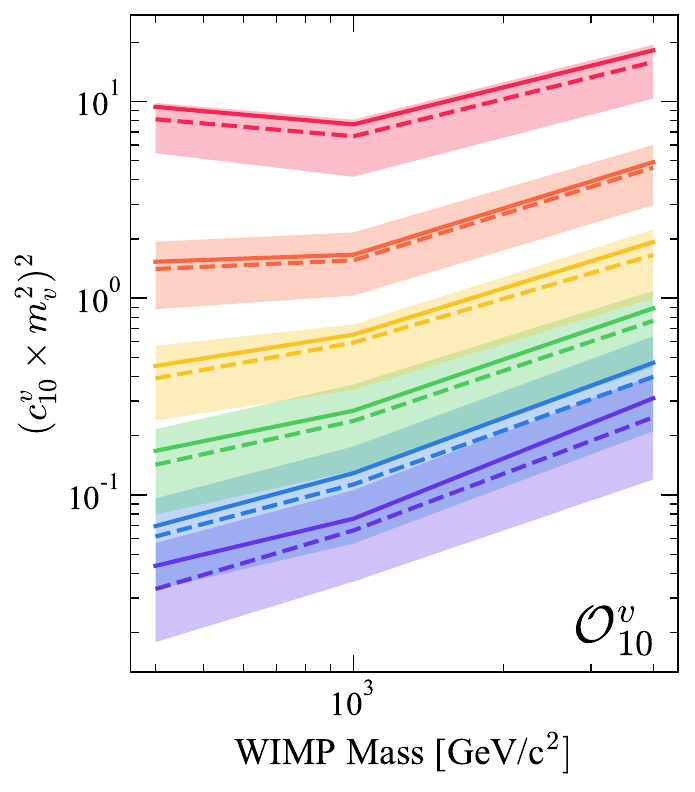}
    \includegraphics[width=0.5\columnwidth]{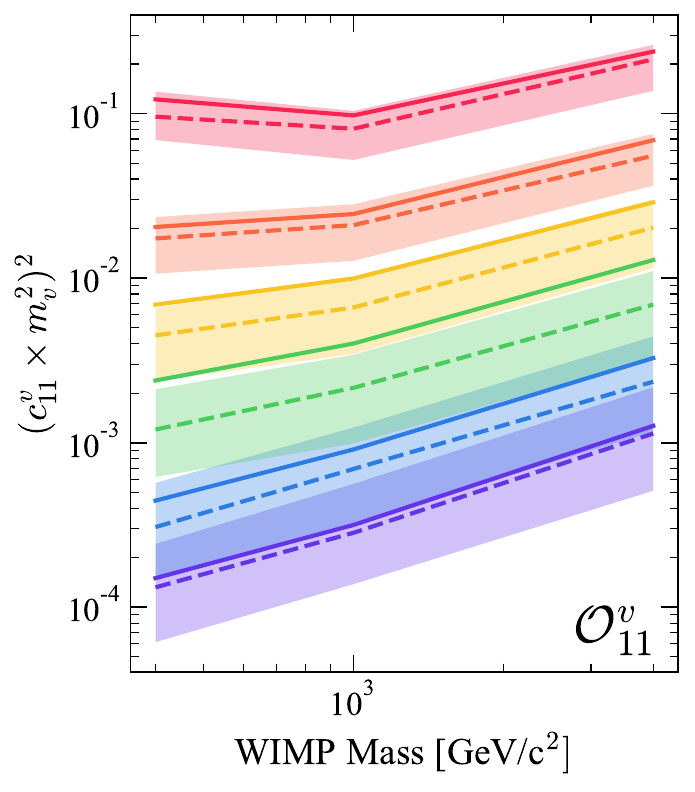}
    \includegraphics[width=0.5\columnwidth]{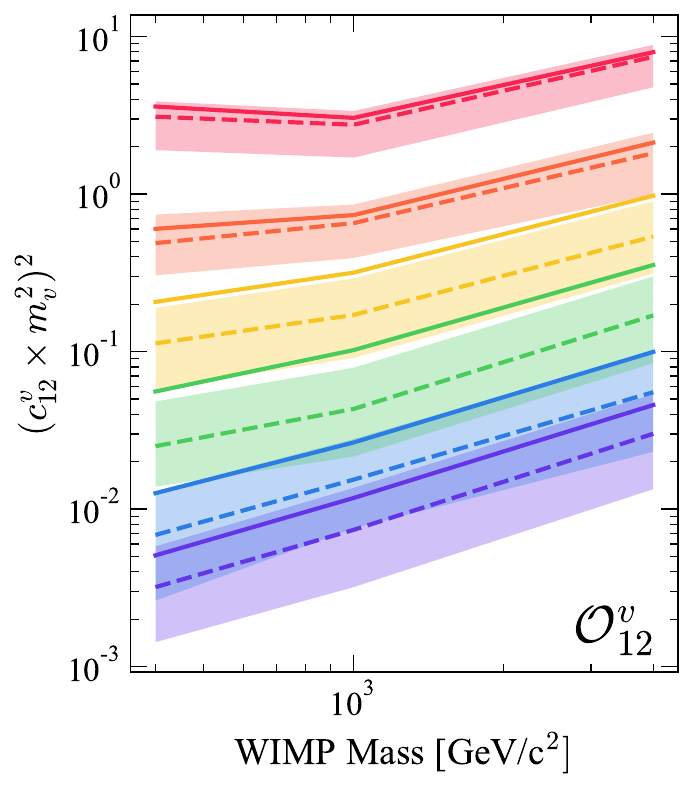}
    \includegraphics[width=0.5\columnwidth]{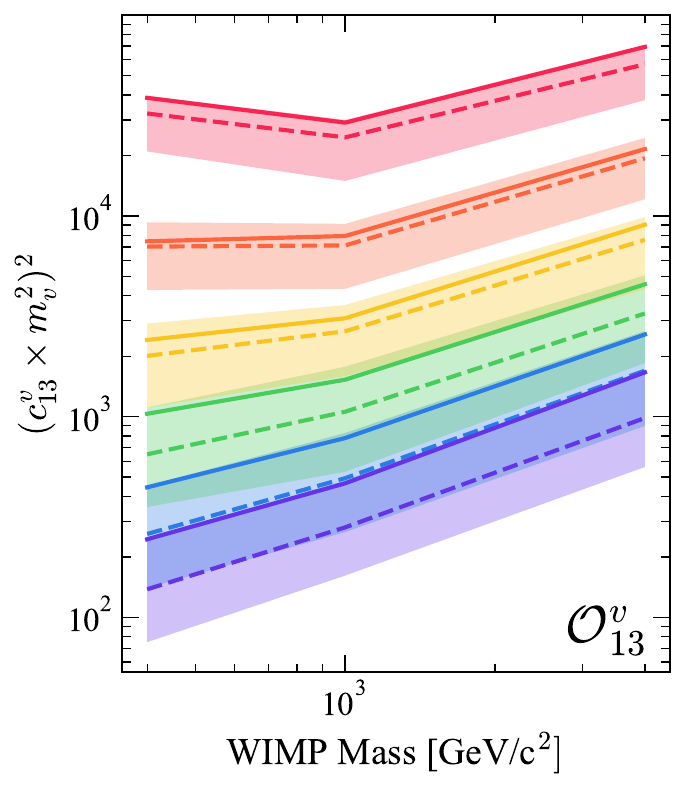}
    \includegraphics[width=0.5\columnwidth]{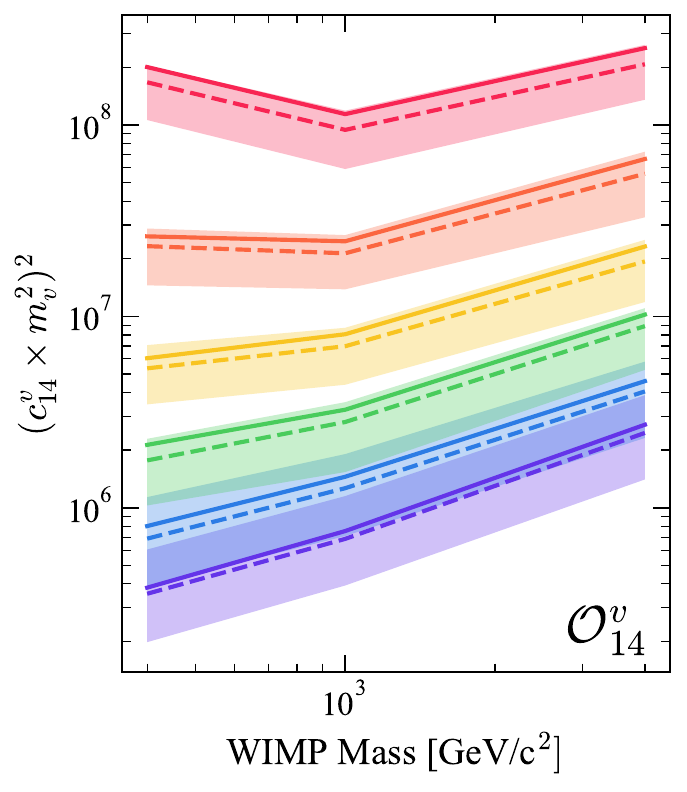}
    \includegraphics[width=0.5\columnwidth]{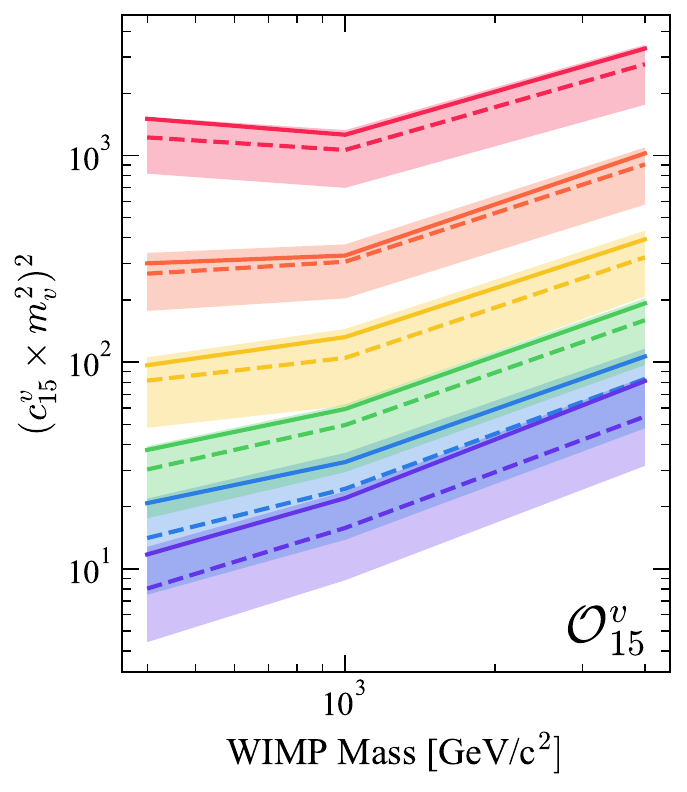}
    \includegraphics[width=0.25\columnwidth,trim=-5cm -20cm 0 3cm]{legend.png}
    \caption{The 90\% confidence limit (solid lines) of the dimensionless isovector WIMP-nucleon couplings for each of the fourteen NREFT interactions.
    The dotted lines show the medians of the sensitivity projection and the shaded bands correspond to the 1$\sigma$ sensitivity band.
    The upper limit is evaluated for WIMP masses of 400~GeV/c$^2$, 1000~GeV/c$^2$, and 4000~GeV/c$^2$, for values of the mass splitting $\delta$ of 0~keV (purple), 50~keV (blue), 100~keV (green), 150~keV (yellow), 200~keV (orange), and 250~keV (red).}
    \label{fig:limits-inelastic-v}
\end{figure*}

Limits on the coupling strengths $c_i^I$ for inelastic interactions are presented in \autoref{fig:limits-inelastic-s} (isoscalar) and \autoref{fig:limits-inelastic-v} (isovector) for values of the mass splitting $\delta$ spanning up to 250~keV and WIMP masses from 400~GeV/c$^2$ to 4~TeV/c$^2$. 
For lighter WIMPs, an increasingly larger fraction of the scattering energy is required to transition to the heavier state. 
Therefore, the inelastic rates for lighter WIMPs increasingly fall below the LZ energy threshold, and so WIMPs lighter than 400~GeV/c$^2$ are not considered. 
The same kinematic suppression leads to weaker limits for larger values of $\delta$ at all WIMP masses \cite{Smith_2001}. 
The observed inelastic upper limits do not drop below 1$\sigma$ of the background-only expectations; therefore, no power constraint is required. 

\section{\label{sec:conclusion}Conclusion}
This article presents the results of a search for a wide range of dark matter scenarios using the first 60~live-day run of the LZ detector.
The study extends the energy window of Ref.~\cite{LZ:SR1WS_2022} to include nuclear recoils of up to around 270~keV$_\text{nr}$ (defined as where the acceptance starts to fall below 90\%), which requires modeling and removal of the $\gamma$-X background.
A frequentist statistical analysis tested the data against 14 WIMP-nucleon operators generated by an NREFT, for both elastic and inelastic scattering, and for isoscalar and isovector couplings to the xenon nucleus.
A total of 56 distinct interactions were tested, corresponding to a set of nuclear recoil spectra that span the entire energy window used in the analysis.
No significant evidence of an excess is observed for any model.

The coupling strengths of all possible elastic (inelastic) interactions between nuclei and dark matter with mass 10--10$^4$~GeV/c$^2$ (400--4000~GeV/c$^2$) are significantly constrained.
LZ provides the strongest measured upper limits for nearly all the models tested.
In particular, models with momentum-suppressed recoil spectra were tightly constrained due to the extended energy window that provided a higher efficiency for the resulting signal events. 
This result paves the way for NREFT interactions to be constrained further by future searches that leverage target nuclei with different isospin properties to xenon \cite{schneck2015dark}. 

\begin{acknowledgments}
The research supporting this work took place in part at SURF in Lead, South Dakota. Funding for this work is supported by the U.S. Department of Energy, Office of Science, Office of High Energy Physics under Contract Numbers DE-AC02-05CH11231, DE-SC0020216, DE-SC0012704, DE-SC0010010, DE-AC02-07CH11359, DE-SC0012161, DE-SC0015910, DE-SC0014223, DE-SC0010813, DE-SC0009999, DE-NA0003180, DE-SC0011702, DE-SC0010072, DE-SC0015708, DE-SC0006605, DE-SC0008475, DE-SC0019193, DE-FG02-10ER46709, UW PRJ82AJ, DE-SC0013542, DE-AC02-76SF00515, DE-SC0018982, DE-SC0019066, DE-SC0015535, DE-SC0019319, DE-AC52-07NA27344, \& DOE-SC0012447.
This research was also supported by U.S. National Science Foundation (NSF); the UKRI’s Science \& Technology Facilities Council under award numbers ST/M003744/1, ST/M003655/1, ST/M003639/1, ST/M003604/1, ST/M003779/1, ST/M003469/1, ST/M003981/1, ST/N000250/1, ST/N000269/1, ST/N000242/1, ST/N000331/1, ST/N000447/1, ST/N000277/1, ST/N000285/1, ST/S000801/1, ST/S000828/1, ST/S000739/1, ST/S000879/1, ST/S000933/1, ST/S000844/1, ST/S000747/1, ST/S000666/1, ST/R003181/1; Portuguese Foundation for Science and Technology (FCT) under award numbers PTDC/FIS-PAR/2831/2020; the Institute for Basic Science, Korea (budget number IBS-R016-D1). 
We acknowledge additional support from the STFC Boulby Underground Laboratory in the U.K., the GridPP~\cite{faulkner2005gridpp,britton2009gridpp} and IRIS Collaborations, in particular at Imperial College London and additional support by the University College London (UCL) Cosmoparticle Initiative. 
We acknowledge additional support from the Center for the Fundamental Physics of the Universe, Brown University. 
K.T. Lesko acknowledges the support of Brasenose College and Oxford University. The LZ Collaboration acknowledges key contributions of Dr. Sidney Cahn, Yale University, in the production of calibration sources. 
This research used resources of the National Energy Research Scientific Computing Center, a DOE Office of Science User Facility supported by the Office of Science of the U.S. Department of Energy under Contract No. DE-AC02-05CH11231. We gratefully acknowledge support from GitLab through its GitLab for Education Program. 
The University of Edinburgh is a charitable body, registered in Scotland, with the registration number SC005336. 
The assistance of SURF and its personnel in providing physical access and general logistical and technical support is acknowledged. We acknowledge the South Dakota Governor's office, the South Dakota Community Foundation, the South Dakota State University Foundation, and the University of South Dakota Foundation for use of xenon.
We also acknowledge the University of Alabama for providing xenon.
For the purpose of open access, the authors have applied a Creative Commons Attribution (CC BY) license to any Author Accepted Manuscript version arising from this submission.
\end{acknowledgments}

\appendix
\section{Comparison of $\mathcal{O}_1$ to Spin-independent\label{ap:1}}
\begin{figure}
    \centering
    \includegraphics[width=0.44\textwidth]{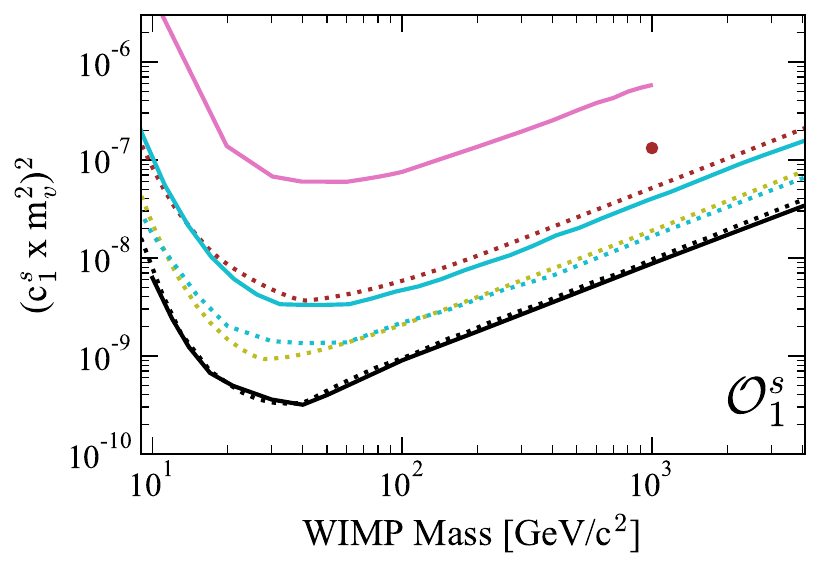}
    \caption{The 90\% confidence limit of the dimensionless isoscalar NREFT $\mathcal{O}_1$ WIMP-nucleon couplings from various LXe direct detection experiments. 
    Solid lines represent the limit on $\mathcal{O}_1$ isoscalar from NREFT analyses: LZ - this work (black), PandaX-II 2019 (blue), XENON100 (magenta) and LUX WS2014-16 (brown point). 
    Dotted lines represent recast SI limits: LZ - this work (black dotted), PandaX-4T 2021 (blue dotted), XENONnT (yellow dotted) and LUX full exposure (brown dotted). 
    The normalizations applied for the recast limits are given in \autoref{tab:isospin_conventions}.}
    \label{fig:xsecComapre}
\end{figure}

The constraints set on the NREFT operators by different LXe TPC direct detection experiments are not always directly comparable.
This is due to variations in the choice of parameters describing the recoil spectra, the use of different conventions and the choice of basis in which the constraints are presented. 
Recasting the results from previous experiments is necessary to draw a comparison to the LZ result.
It is possible to account for some of these differences: isospin representation and dimensions of the result. 
However, differences arising from the choice of the nuclear shell model and dark matter velocity parameters result in non-linear scaling terms and, therefore, can not be easily accounted for. 
As the updated GCN5082 values do not alter the form factor associated with $\mathcal{O}_1$, this operator facilitates a method to compare different experimental results. 
The variations in velocity parameters are not considered here.
Additionally, the isoscalar representation of $\mathcal{O}_1$ allows for comparison against established SI cross-sections.
\autoref{fig:xsecComapre} shows the recasting of various LXe TPC SI and $\mathcal{O}_1^s$, indicating consistency between the two limits produced by the LZ experiment.
\autoref{tab:isospin_conventions} outlines the conversion for each limit.

\begin{table*}[t]
    \centering
   \caption{LXe direct detection experiments that have reported SI and $\mathcal{O}_{1}^{s}$ limits.
   For NREFT limits, the isospin basis used is given for each experiment. 
    The dimension in which the limit was reported is given for each as well as the conversion used to recast the limit in \autoref{fig:xsecComapre}.}
    \begin{ruledtabular}
    \begin{tabular}{p{0.25\linewidth} p{0.21\linewidth} p{0.2\linewidth} p{0.22\linewidth}}
        Experiment & Basis   &Limit Type & Conversion in plot \\ \hline
        Xenon100: 2017 NREFT \cite{Xenon100:EFT_2017}   & \makecell{$c^s=\frac{1}{2}(c^p + c^n)$ \\ $c^v=\frac{1}{2}(c^p - c^n)$} & $(c^{s}_i \times m_v^2)^2$    & None  \\
        LUX: WS2014-16 NREFT \cite{LUX:EFTR4_2021}                 & \makecell{$c^s=(c^p + c^n)$ \\ $c^v=(c^p - c^n)$}      & $(c^{s}_i \times m_v^2)^2$     & $\frac{1}{4}$\\
        PandaX-II: SD EFT \cite{PandaX2:SD_EFT_2019}& \makecell{$c^s=\frac{1}{2}(c^p + c^n)$ \\ $c^v=\frac{1}{2}(c^p - c^n)$}  & $d_5^{s/v}$ [$\frac{1}{m_v^2}$]      &  $(d_5^s)^2$ \\
        LZ NREFT (This analysis) & \makecell{$c^s=\frac{1}{2}(c^p + c^n)$ \\ $c^v=\frac{1}{2}(c^p - c^n)$} & $(c^{s/v}_i \times m_v^2)^2$    & None    \\
        NRET Theory paper \cite{Anand:MathematicaEFT}              & \makecell{$c^s=\frac{1}{2}(c^p + c^n)$ \\ $c^v=\frac{1}{2}(c^p - c^n)$}   & N/A & N/A    \\
        LUX: Combined 2017 SI \cite{LUX:SI_complete2017}& N/A & $\sigma_{SI}^N$ & $ \sigma^N_{SI} \frac{\pi \cdot m_v^4}{(\frac{(\hbar c)}{\rm{GeV}})^2\mu_N^2}$ \\
        PandaX-4T: 2021 SI \cite{PandaX4T:SI2023} & N/A & $\sigma^N_{SI}$ & $\sigma^N_{SI} \frac{\pi \cdot m_v^4}{(\frac{(\hbar c)}{\rm{GeV}})^2\mu_N^2}$ \\
        LZ: 2023 SI \cite{LZ:SR1WS_2022} & N/A & $\sigma^N_{SI}$ & $\sigma^N_{SI} \frac{\pi \cdot m_v^4}{(\frac{(\hbar c)}{\rm{GeV}})^2\mu_N^2}$ \\
        XENONnT: 2023 SI \cite{XenonNT:WS_2023} & N/A & $\sigma^N_{SI}$ & $\sigma^N_{SI} \frac{\pi \cdot m_v^4}{(\frac{(\hbar c)}{\rm{GeV}})^2\mu_N^2}$ \\
    \end{tabular}
    \end{ruledtabular}
    \label{tab:isospin_conventions} 
\end{table*}

\newpage
\section{Data Release\label{ap:2}}
Data from selected plots in this paper can be accessed at: \href{https://www.hepdata.net/record/ins2729878}{https://www.hepdata.net/record/ins2729878}

\begin{itemize}
    \item[--] \autoref{fig:acceptances}:
   points representing the total efficiency curve for this analysis (black line).
    \item[--] \autoref{fig:data-ROI}: points in S1-S2 space representing the data used in the WIMP search (black points).
    \item[--] \cref{fig:limits-elastic-s,fig:limits-elastic-v,fig:limits-inelastic-s,fig:limits-inelastic-v}: Points representing the observed 90\% confidence level upper limits, together with the median, $\pm1\sigma$, and $\pm2\sigma$ expected sensitivities.
\end{itemize}

\newpage
\bibliography{biblog}

\end{document}